\documentclass{JHEP3}
\usepackage{amsmath,amsfonts,bbm,euscript, amssymb}
\usepackage{cite}
\usepackage{epsfig}
\usepackage{graphicx}
\usepackage{psfrag}
\def\al{\alpha}
\def\om{\omega}

\def\be{\begin{equation}}
\def\ee{\end{equation}}
\def\baray{\begin{eqnarray}}
\def\earay{\end{eqnarray}}
\def\ba{\begin{eqnarray}}
\def\ea{\end{eqnarray}}

\title{The Boundedness of Euclidean Gravity and the Wavefunction of the Universe
}
\author{Sash Sarangi and S.-H. Henry Tye \\
 Laboratory for Elementary Particle Physics, Cornell
University, Ithaca, NY 14853 \\
E-mail:\email{sash@lepp.cornell.edu},
\email{tye@lepp.cornell.edu}}

\abstract{When the semi-positive cosmological constant is dynamical, the naive Euclidean Einstein 
action is unbounded from below and the Hartle-Hawking wavefunction of the universe is not normalizable. With the inclusion of back-reaction (a crucial point), the presence of the metric 
perturbative modes (as well as matter fields) as a radiation term is introduced by quantum fluctuation. They act as the environment (that is, to be integrated or traced out), and 
introduce a correction term that provides a bound to the Euclidean action. As a result,  the improved wavefunction is normalizable. That is, decoherence plays an essential role in the  consistency of quantum gravity. In the spontaneous creation of the universe, this improved 
wavefunction allows one to compare the tunneling probabilities from absolute nothing (i.e., not even 
classical spacetime) to various vacua (with different large spatial dimensions and different low 
energy spectra) in the stringy cosmic landscape.
}

\begin{document}

\tableofcontents

\section{Introduction}

In the probing of the origin of our universe, a particularly attractive idea is the tunneling from 
absolute nothing (here, nothing means not even classical spacetime) \cite{Vilenkin:1982de}, or 
equivalently, the Hartle-Hawking (HH) no-boundary wave function of the universe \cite{Hartle:1983ai}.
The basic idea is illustrated in Figure 1. 
\begin{figure}
\begin{center}
\epsfig{file=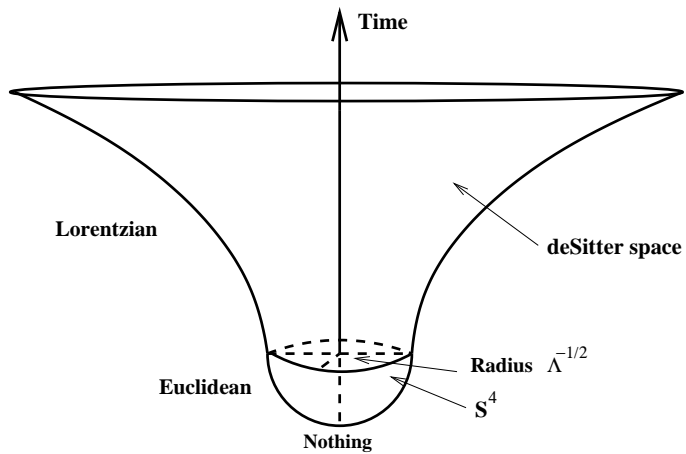, width=9cm}
\vspace{0.1in}
\caption{Tunneling from absolute nothing to a deSitter or Inflationary Universe.}
\label{fig1}
\end{center}
\end{figure}
It is strongly believed that, in superstring theory, there is a vast number of stable and 
meta-stable vacua, with up to 9 or 10 large spatial dimensions. This is the cosmic landscape.
If we can reliably calculate the tunneling probability from absolute nothing 
%(that is, no classical spacetime)
to any point in this vast landscape, one may argue that the origin of our universe should be 
the point in the landscape with the largest tunneling probability. 
This tunneling probability is given by (the absolute square of) the wavefunction of the universe. However, the HH wavefunction is not normalizable. 
Furthermore, it gives an answer that contradicts observations. It is well-known that the 
problem lies in the infrared/macroscopic limit, so quantum/stringy corrections will not be helpful. 
In Ref\cite{Firouzjahi:2004mx}, we conjectured that a normalizable wavefunction results if 
decoherence effect is included. One can then apply this improved wavefunction to find the preferred vacuum in the cosmic landscape (the one emerging from tunneling from nothing with the largest tunneling probability), which then evolves to today's universe. We call this proposal the Selection of the Original Universe Principle (SOUP); that is, today's universe must lie along the road that starts with the 
original preferred vacuum, arguably the 4-dimensional inflationary universe supported by observations. In this paper, we investigate more carefully the decoherence effect on the 
tunneling probability/wavefunction. As expected, the result supports the basic underpinning of SOUP, though the details are somewhat more involved. 
 
Consider the simple Einstein theory in 4 dimensions with a positive cosmological constant :
\ba
S_{E} = - \frac{1}{16 \pi G}\int d^4 x \sqrt{-g} \left( R - 2\Lambda\right)
\ea
The tunneling probability from nothing to a closed deSitter (or inflationary) universe with 
cosmological constant $\Lambda$ is given by
\ba
\label{EactionG}
\Gamma = |\Psi|^2\simeq e^{-S_E}  \quad \quad S_E=- \frac{3 \pi}{G\Lambda} 
\ea
where $S_E$ is the Euclidean action of the $S^4$ instanton \cite{Gibbons:1978ac}.
Suppose $\Lambda$ is dynamical, as in a model with 4-form field strengths \cite{Brown:1987dd}. 
The tunneling probability $\Gamma$ seems to allow us 
to pick out the universe with the largest probability. However, the Euclidean action $S_E$ is 
unbounded from below as $\Lambda \rightarrow 0$, so $\Gamma \to \infty$.
This has at least 2 obvious problems :

(1) The wavefunction $\Psi \simeq e^{3 \pi/2G\Lambda}$ is not normalizable. 
%Unitarity can be satisfied if $I_E$ is bounded from below. If the above formula is correct, 
%there is no normalization that can satisfy unitarity. 
Furthermore, one can easily show that there are other topological instantons with even 
more negative Euclidean action and so larger (actually, infinitely larger)  tunneling 
probability \cite{Coleman:1988tj}. This implies the presence of an inconsistency.

(2) Phenomenologically, since $\Gamma \rightarrow \infty$ as  $\Lambda \rightarrow 0$, 
it will imply the preference of tunneling to a flat universe with zero $\Lambda$, which 
contradicts the big bang history of our universe. 
Since the size of the universe, the cosmic scale factor $a, \sim 1/\sqrt{\Lambda}$, this will imply 
that the tunneling is to a universe of super-macroscopic size, contradicting one's intuition
that tunneling is a quantum process and so should be microscopic.
 
This issue is an outstanding problem since the early 1980s. 
The inconsistency has prevented the proper application of the whole idea.  
Possible resolution to this problem has been suggested (see 
\cite{Linde:1998gs} and references therein):
\begin{itemize}
\item
Instead of the usual $t \rightarrow - i \tau$, 
one may choose instead to rotate time to Euclidean time via 
$t \rightarrow i \tau$ \cite{Linde:1984mx}. This may work for pure gravity, but the 
inclusion of a scalar field (e.g., as required for inflation) 
leads to catastrophic consequences, since 
the scalar field theory becomes unbounded from below \cite{Rubakov:1984bh}. 
\item
One may argue that this problem will be corrected by quantum 
corrections or string theory corrections. However, it is easy 
to see that this is unlikely to be the case. Note that the 
problem occurs for small $\Lambda$, or large universe, since 
the cosmic scale factor $a \sim 1/\sqrt{\Lambda}$.
So this is more like an infrared or macroscopic problem than an ultra-violet problem. 
In fact, one can easily see that the loop correction \cite{Hawking:1976ja,Gibbons:1978ji}
does not solve the problem. Also, recent work \cite{Ooguri:2005vr}, where the exact HH wavefunction
is obtained in topological string theory, it seems that 
$\Lambda \to 0$ is again preferred, just like the original HH wavefunction.

\item This last property, namely, that the unboundedness problem 
becomes acute when the deSitter universe becomes large, or 
macroscopic (actually super-macroscopic), naturally suggests that 
the resolution should lie in decoherence \cite{Firouzjahi:2004mx}.
\end{itemize}

In Ref.\cite{Firouzjahi:2004mx}, we argue that the mini-superspace formulation is inadequate, 
and we propose that the inclusion of decoherence effects due to other modes provides a lower bound to $S_E$. We then speculate how the improved 
wavefunction may be used to select the stringy vacuum with the largest tunneling probability
from absolute nothing.
In this paper, we shall show that decoherence indeed provides a bound to the Euclidean gravity
action, though the formula for the tunneling probability may be more involved.  
Here, we shall focus on the pure 4-D Einstein gravity case. Generalization to more dimensions is straightforward and will be briefly discussed.

%add this
The basic idea of the approach is widely used in physics. Given any complicated problem, we usually
follow only a limited set of degrees of freedom, called the system. The remaining degrees 
of freedom, called the environment, are either ignored, or, in a better approximation, integrated out. 
In decoherence, integrating out the environment can cause the quantum system to behave like
a classical system. In effective field theory in particle physics, the massive modes are integrated 
out to produce higher 
dimensional operators (interaction terms) for the light modes. A famous example is the integrating
out of the W and Z bosons in the electroweak theory that yields the 4-Fermi weak interactions. Another example is the integrating out of the hidden sector in supergravity phenomenology. (Note that this has nothing to do with loop corrections.)
In the Wilson approach in quantum field theory, where high momentum modes are integrated out, and in the above cases, it is clear that the physics may crucially depend on the effects coming from integrating out the unobserved modes; that is, they cannot be ignored. 
%
%Integrating out the environment provides an effective bound to the resulting theory can be seen in a simple example. 
%
%For an example more closely related to the problem at hand, consider a somewhat artificial model of 2 scalar fields $\phi$ and $\eta$, with an effective potential $V= -m^2\phi^2/2 + M^2\eta^2/2 + \lambda \phi^2 \eta$. Here $\phi$ may have a tachyonic mass. Either $\eta$ has a large tachyonic mass (technically, we should treat it as an auxiliary field without a kinetic term) or $\lambda$ is imaginary.  At low energy, if we ignore $\eta$ completely, we find that the effective potential $V(\phi)$ is unbounded from below. However, integrating out the heavy mode $\eta$ introduces an effective interaction term $|\lambda|^2 \phi^4/M^2$, which provides a lower bound to the resulting effective potential $V(\phi)$. 
%
%The situation in Euclidean gravity is quite similar.
Of course, loop corrections can also induce dynamics/interactions that are not present in the tree level, as for example in the Coleman-Weinberg model, in light-light scattering, and in the running of couplings. In QCD, the running of the coupling emerges from the renormalization group improved quantum corrections, not just a normal 1-loop effect.
%however, this is not our main concern here.
We argue that this is also the situation in the study of the wavefunction of the universe. One may consider our result as a back-reaction improved quantum correction, not just a normal 1-loop effect.

The basic idea applied to tunneling is quite simple. It is well-known that the quantum tunneling of a particle with mass $M$, or the system, is suppressed if it interacts with an environment.
Consider a particle at $q=0$ in the potential $V(q)$ as shown in Figure 2.
\begin{figure}
\begin{center}
\epsfig{file=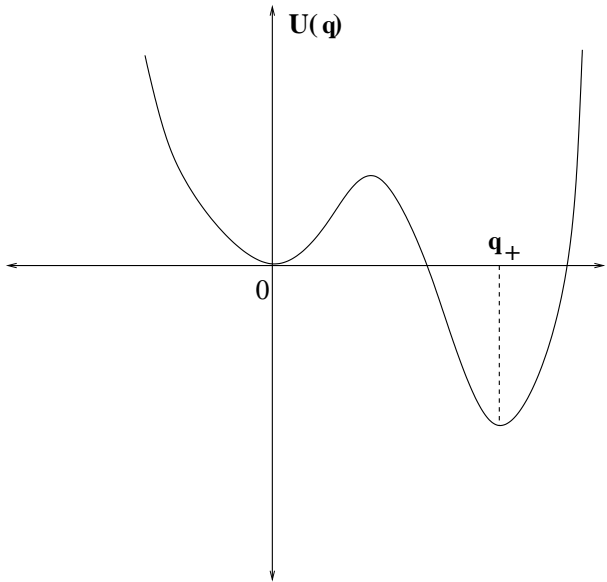, width=8cm}
\vspace{0.1in}
\caption{Potential $V(q)$ with a metastable minimum at
$q=0$, and a stable minimum at $q = q_{+}$.}
\label{fig2}
\end{center}
\end{figure}
%The WKB approximation is good provided that the height of the barrier is larger than 
%$\om_0$, where $$M \om_0^2 =  \frac{\partial^2 V}{\partial q^2}|_{q=0}$$
In the WKB approximation, its tunneling rate is given by 
\ba
\Gamma &\simeq&  \exp (-S_0) \nonumber \\
S_0 &=& \int^{q_0}_0 \sqrt{2MV(q)} dq
\ea
where $S_0$ is the Euclidean action of the bounce, i.e., the instanton 
solution \cite{Coleman:1977py}.
Note that, for $V(q)$ bounded from below, $S_0$ is bounded 
from below, as required by consistency.

In a more realistic situation, the particle interacts with a set of other particles, or modes, 
say $x_{\al}$. However, we are only interested in the quantum status of $q$, so these other modes 
are integrated out in the path integral, or traced over in the density matrix formulation. 
They provide the environment. Their presence typically introduces a frictional force to 
the evolution of $q$.  It was shown by Sethna \cite{Sethna:1981dr} and by Caldeira and Leggett \cite{Caldeira:1982uj}, that the bounce $S_0$ increases to
$S \simeq S_0 \left( 1 + {\hat \eta}\right)$
where $\hat \eta > 0$ is proportional to the coefficient of friction (see Appendix A for details). 
That is, the interaction with the environment suppresses the tunneling rate. (This suppression takes place even if no friction is generated.) One may understand this result in a 
number of (equivalent) ways :
\begin{itemize}
\item 
As a quantum system, 
\ba
S &=& \int  \sqrt{2MV(q,x_{\al})} ds \nonumber \\
ds^2 &=& dq^2 + \sum \frac{m_{\al}}{M} d x_{\al}^2
\ea
That is, the increase in $S$ is due entirely to the longer path 
length in the many dimensional $(q, x_{\al})$ space.
\item 
The interaction of $q$ with the $x_{\al}$ interferes with its 
attempt in tunneling. One may view the interaction with $x_{\al}$ 
as attempts to observe $q$. Repeated measurements of $q$ or 
repeated attempts of measuring $q$ suppresses the tunneling rate.
This is analogous to the Zeno or Watch Pot effect.
\item
The interaction of $q$ with the environment $x_{\al}$ diminishes 
the quantum coherence. As a consequence, the system behaves more 
like a classical system than like a quantum system. Since tunneling is a 
pure quantum phenomenon, it should be suppressed as the system becomes 
more macroscopic/classical. We shall refer to this as decoherence, by which we mean
the process where a quantum system behaves more classically (i.e., less quantum)
via its interactions with the environment \cite{zeh}.

\end{itemize}

Here, we study this effect in quantum gravity, in the tunneling 
from nothing scenario. In this case, the cosmic scale factor $a$ 
plays the role of the system, while the metric fluctuations around $a$ (and any matter field modes)
play the role of $x_{\al}$, i.e., the environment. Figure 3 illustrates this situation. 
\begin{figure}
\begin{center}
\epsfig{file=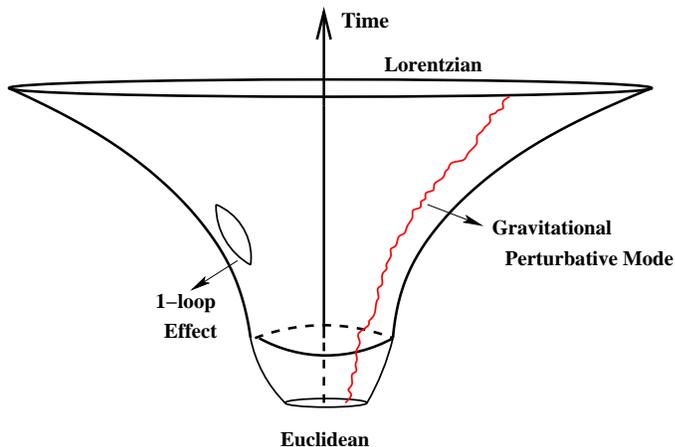, width=9cm}
\vspace{0.1in}
\caption{Typical tunneling involves metric and matter excitations as well. Loop effects may also be included, though they are not important for our discussions.}
\label{fig3}
\end{center}
\end{figure}
Since we measure only $a$, the metric fluctuations are integrated out 
in the path integral (or traced out in the density matrix).  
As expected, we shall show that their presence suppresses the tunneling
from nothing to a deSitter universe.
As expected, the decoherence effect is negligible for large cosmological constant 
(small size universe), but becomes
increasingly important as $\Lambda \to 0$, as 
it suppresses the tunneling probability when $a = 1/\sqrt{\Lambda/3}$ becomes macroscopic.
In contrast to a normal quantum system, the inclusion of the environment is of 
fundamental importance in quantum gravity, since the corrected Euclidean 
action is now bounded from below. 

The coupling of the metric fluctuations and scalar fields to $a(\tau)$ are more complicated than that of 
the above quantum system. Each metric perturbative mode behaves like a simple harmonic 
oscillator but with time-dependent (or $a$-dependent) mass and frequency. However, the real subtlety of the calculation comes in another way. If we treat the metric fluctuation modes as pure perturbations, 
we shall get nothing except the loop correction, a known result in Euclidean gravity. 
This is not hard to see. The Euclidean action in minisuperspace (that is, keeping only the cosmic scale factor) is given by
\ba
S_{E} = \frac{1}{2}\int d\tau \left( -a\dot{a}^2 - a + \lambda a^3 \right)
\ea
where a rescaling has rendered $\tau$, $a(\tau)$, the Hubble constant $H=1/\sqrt{\lambda}$ and $\lambda=2G\Lambda/9 \pi$ dimensionless.
For $S^4$, $Ha(\tau)= \sin (H \tau)$, where south pole (north pole) corresponds to $H \tau= 0$ ($\pi$). This path gives the Euclidean action (\ref{EactionG}). For a fluctuation mode $f$ that satisfies the classical equation, the Euclidean classical action can be written as a surface term
$S_E \simeq a^3 f {\dot f} {\large |}^{\pi/H}_0=0$, 
since $a(\tau)=0$ at the two poles. As a result, no decoherence term is generated. (Including an additional boundary term makes no difference.)
However, we find that the back-reaction is crucial for getting the correct answer. Instead of using the unperturbed $a(\tau)$ given above for the $S^4$ geometry, we leave it arbitrary during the tracing 
out of the fluctuation modes. 
In the path-integral formalism, one starts with the path integral that includes the scale factor $a$  as well as the perturbations around $a$. The following, for example, 
shows the inclusion of the metric tensor perturbations $t_n$ 
\ba
\label{path}
Z = \int D[a]\prod_{n}\int_{t_n^i}^{t_n^f}D[t_n]e^{-S_E} \exp\left( 
 -\frac{1}{2}\int_{-T/2}^{T/2} d\tau a^3 \sum_{n} [\dot{t_n}^2 + \omega_n(a)^2
t_n^2]  \right)
\ea
Tracing out the perturbations, we have 
\ba
\label{trace}
Tr[Z] = \prod_{n} \int dt_n^i \int dt_n^f \delta(t_n^i - t_n^f)~ Z
\ea
This results in a new term in the modified action  
\ba
S_{E,dC} \simeq \frac{1}{2}\int d\tau \left( -a\dot{a}^2 - a + \lambda a^3 + \frac{\nu}{\lambda^2 a}\right)
\ea
where the last term, coming from integrating out the perturbative modes, behaves like ordinary radiation. $\nu$ is a constant that measures the number of perturbative modes.  We then solve for $a(\tau)$ and obtain the corrected Euclidean action $S_{E,dC}$ in the saddle-point approximation.
It turns out that the $\nu$ term modifies the shape of the $S^4$ instanton to barrel-like, as shown in Figure 4. Since the effect is perturbative in nature, we expect the barrel to have the same topology as $S^4$. We see that, due to the back-reaction, $a(\tau)$ does not vanish at the two ends. However, the contribution of the end plates of the barrel to $S_{E,dC}$ happens to be zero.
 
After tracing over the metric fluctuations in this way, the rate 
of tunneling from nothing (i.e., no classical spacetime) to a 
deSitter universe (much like the inflationary universe) is now
given by $\Gamma \simeq \exp(F)= \exp(-S_{E,dC})$, where
\ba
\label{FS}
- F = S_{E,dC} = S_{E,0}+ D \simeq -\frac{3 \pi}{G\Lambda} + c \left(\frac{3 \pi}{G\Lambda}\right)^2
\ea
where $D$ is the decoherence term and $c$ depends on the cut-off.
In string theory, that cut-off is naturally provided by the string scale.
Note that $S_{E,dC}$ is now bounded from below. See Figure 5.
Here, $c$ in string theory also depends on the string spectrum, so 
$c$ should be calculated for each vacuum. 
%We shall call the second term the decoherence term.
We find that 
 \ba
 c = \frac{3}{2} n_{dof} \left( \frac{2 G}{3\pi l_{s}^2} \right)^2
 \ea
where $l_s$ is the string scale ($\alpha^{\prime} = (l_s/2 \pi)^2$) and 
$n_{dof}$ is the number of light degrees of freedom included in the environment.
For the pure gravity case, we have $n_{dof}=2$ for the two tensor modes.
For small $\Lambda$, below a critical value, the tunneling is actually totally suppressed.
\begin{figure}
\begin{center}
\epsfig{file=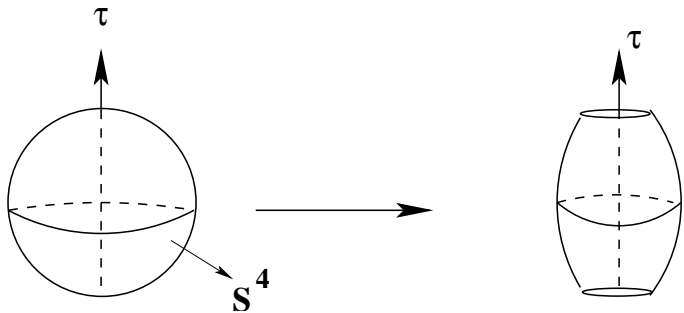, width=9cm}
\vspace{0.1in}
\caption{Modification of the $S^4$ instanton due to the presence of decoherence, 
i.e., the inclusion of the environment.}
\label{fig4}
\end{center}
\end{figure} 

This is quite understandable. A precise $S^4$ spherical geometry leads to the pure de Sitter space, which by definition excludes any radiation. If we do not allow back-reaction, then the system $a(\tau)$
actually cannot feel the presence of the environemnt. Allowing back-reaction, the quantum fluctuation
during the spontaneous creation of the universe generates some radiation (even though no 
radiation is introduced classically). They act as the environment.
The presence of this environment modifies the geometry in a way so that tunneling is to a universe with both a cosmological constant and some radiation, with a suppressed tunneling rate. 
The presence of the radiation generates the decoherence term. Note that, in integrating out the perturbative modes, the case with zero amplitude for the perturbative modes is also included.
As we see, the amount of radiation is proportional to $1/\Lambda^{2}$, so there is more radiation in a larger universe. 

What we consider fundamental is that the decoherence effect 
actually provides the Euclidean action $S_E$ of pure gravity with a lower bound.
Since the metric fluctuation contribution to $S_E$ cannot be 
turned off, they must be included in the evaluation of the tunneling rate. 
For usual quantum system, $S_0$ is bounded from below. So 
one may view this friction/environment/decoherence effect as 
a correction, albeit it may be very big. In quantum gravity, 
this effect resolves the unboundedness problem.
That the quantum fluctuation provides a natural source to cure the boundedness problem 
implies that quantum gravity is actually self-consistent in the macroscopic regime. 

\begin{figure}
\begin{center}
\epsfig{file=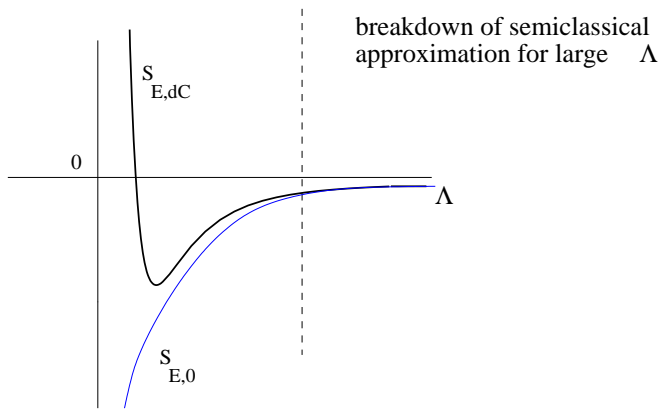, width=9cm}
\vspace{0.1in}
\caption{$S_{E,dC}$ and $S_{E,0}$ as functions of the cosmological constant $\Lambda$.
We see that, with the inclusion of the decoherence effect, the Euclidean action for the
$S^4$ instanton  is now bounded from below. For large $\Lambda$, the semi-classical approximation breaks down. For small $\Lambda$, tunneling is actually completely suppressed.
}
\label{fig5}
\end{center}
\end{figure}
This effect renders 
the system more macroscopic and so less quantum. That is, as $a$ 
becomes large, its interaction with the metric fluctuations and matter fields should 
suppress the tunneling rate. Indeed, we show that the inclusion of 
the metric fluctuation decreases the tunneling rate, as expected.
Note that the decoherence term is not the usual perturbative 
quantum correction. For large $\Lambda$, where quantum correction 
is expected to be large, the decoherence term actually becomes negligible.
One can include the quantum corrections; however, they do not change 
the qualitative behavior for moderate values of $\Lambda$, i.e., $G \Lambda <<1$. 

Let us gain some idea of the magnitude of $F$. Following Eq.(\ref{FS}), we find the value of $\Lambda$ with maximum tunneling probability is
\ba
\Lambda_{max} = \frac{4 n_{dof} G}{\pi l_s^4}  \to F_{max}=\frac{3}{2n_{dof}} (2 \pi M_{Pl} l_s)^4
\ea
For the string scale a few orders of magnitude smaller than $M_{Pl}$, we can easily have $F \simeq 10^{10}$ or larger. On the other hand, the critical value of $\Lambda$ is $\Lambda_c=\Lambda_{max}/2$. For $\Lambda < \Lambda_c$, the barrel-shaped instanton is destroyed. At $\Lambda_c$,
$F(\Lambda_c)=0$, so the tunneling probability at $\Lambda$ close to $\Lambda_c$ is already negligibly small compared to that at $\Lambda_{max}$.

Having resolved the outstanding problem mentioned above, one 
may then apply this consistent tunneling approach to select 
a preferred vacuum in the cosmic landscape in string theory.
It is straightforward to generalize Eq.(\ref{FS}) to arbitrary spatial dimensions. In particular, for ten dimensional spacetime, we have
\ba
\label{FS10}
- F = S_{E,dC} \simeq S_{E,10}  + c \left(\frac{V_{10}}{l_s^{10}}\right)
\ea
where $S_{E,10}$ is the 10-D Euclidean action determined in mini-superspace and 
$V_{10}$ is the 10-dimensional volume of the instanton. 
Note that (see Ref\cite{Firouzjahi:2004mx}) $S_{E,10}$ reduces to $-{3 \pi}/{G\Lambda}$
for a vacuum where the extra 6 dimensions are compactified.
For each vacuum, $c$ depends on 
the spectrum. It may also depend on the compactification and dilaton moduli.
Knowing these properties of each vacuum allows one to calculate its 
tunneling probability from nothing. One may estimate the size of $c$ by comparing to the 4-D case:
\ba
c \simeq \frac{n_{dof} } {\pi M_{Pl}^2 l_s^2g_s^2}
\ea
with $8\pi G= 1/M_{Pl}^2$ and $g_s$ is the string coupling. Since the string scale is expected to be a few orders of magnitude below the Planck scale, we expect $c$ to be a small number.
In \cite{Firouzjahi:2004mx}, we consider the suppression
of tunnelling in the context of the spontaneous creation of the 
universe. The suppression happens due to the effect of both
the gravitational and matter perturbations on the bounce solution
to the Euclidean Einstein equations.  The general idea, loosely
speaking,  is to separate the ``universe'' into a ``system'' (the pure
gravitational bounce) and the ``environment'' (the perturbations). 
One is interested in measuring properties of the system and in order 
to do so one simply traces out the environmental degrees of freedom. The effect of 
the environment can be significant and such effects have been studied in 
various systems. In the previous work \cite{Firouzjahi:2004mx} 
we estimated this effect  on the tunnelling probability 
of the universe by considering the unperturbed deSitter space as the 
``system'' and the ``environment'' consists of the metric perturbations and matter fields.
We see that the qualitative features are robust, though the details are somewhat more 
involved. (There we used $V_9$ instead of $V_{10}$ for Eq.(\ref{FS10})). 
For large $\Lambda$, the semi-classical approximation 
breaks down. For small $\Lambda$, additional decoherence effect may be important. 
(Here we have calculated only the leading order.)
Fortunately, the range of $\Lambda$ where the tunneling probability is largest seem to lie
in the region where the approximation is most reliable; and we are interested in vacua 
with large tunneling probabilities. Since the tunneling probability drops off rapidly as 
$\Lambda \to 0$ and $\Lambda \to \infty$, their precise values are not 
as important to us. With some luck, we may use the above formula to locate the preferred 
set of vacua in the cosmic string landscape. Note that $c$ depends on the spectrum at each
point in the landscape. 

As one can see in Figure 5, intermediate values of $\Lambda$, much like the inflationary universe that describes the history of our universe \cite{Guth:1981zm} seems to be preferred.
As in Ref\cite{Firouzjahi:2004mx}, we find phenomenologically that 10-dimensional deSitter-like 
vacua (with $F \simeq 10^9$ for instantons $S^{10}$, $S^5 \times S^5$, $S^4 \times S^3 \times S^3$) are not 
preferred while supersymmetric vacua in any dimension have essentially zero tunneling probability. Also, tunneling to vacua very much like our today's universe (with a very small dark energy) seems to be severely suppressed. Tunneling to a universe with quantum foam seems not preferred either. Among the known vacua, the preferred ones are the 4-D brane inflation \cite{Dvali:1998pa} as realized in a realistic string model \cite{Kachru:2003sx}, very much like the inflationary universe that our universe has gone through (with $F \sim 10^{16}$). Although the details of the decoherence term obtained here is somewhat different from that used in Ref\cite{Firouzjahi:2004mx}, we see that the qualitative features summarized there remain true.

Decoherence and related issues in quantum/Euclidean gravity have been studied earlier 
\cite{Halliwell:1985eu,Kiefer:1987ft,Kiefer:1989ud,Halliwell:1989vw,Kiefer:1992cn}, where they are
mostly concerned with the evolution of the inflationary universe. Here, we study decoherence in the 
quantum tunneling in gravity. To be self-contained, we shall review some of the relevant formalisms developed there.

In Sec. 2, we review the tunneling to closed deSitter space and perturbative modes around 
the cosmic scale factor $a(\tau)$.
In Sec. 3, we evaluate the contribution due to these modes to the effective Euclidean action. Here we find that no decoherence term is generated in the $S^4$ background. To see why back-reaction is important, the effect is calculated in the  background of a ``squashed'' $S^4$ geometry. The reader may choose to skip this section. 
In Sec. 4, we study the modified bounce including back reaction. We derive the effective Euclidean action after integrating out the metric perturbation with an arbitrary background $a(\tau)$.
In Sec. 5, We find the modified bounce solution from the effective Euclidean action. This allows 
us to obtain the improved wavefunction and tunneling probability from nothing. 
In Sec. 6, we show that the new term in the improved Euclidean action behaves like ordinary radiation.
We see also how the new term appears in the Wheeler-DeWitt equation and how it influences the tunneling amplitude. 
In Sec. 7, we discuss the implication of the main result and its connection to the proposal of 
Ref\cite{Firouzjahi:2004mx}.
Sec. 8 contains some discussions and Sec. 9 contains a summary and further remarks. Some of the details are relegated to appendices.
For the sake of completeness, a number of known results are reviewed extensively.

\section{Setup}

\subsection{Notations}

We shall render various quantities dimensionless for the ease of calculation.
The conventions we follow are those of \cite{Hartle:1983ai}. The 
Euclidean action is defined as
\ba
S_{E} =-\frac{1}{16 \pi G}\int d^4 x \sqrt{-g} \left( R - 2\Lambda\right)
\ea
The Euclidean metric is given by
\ba
ds^2 = \sigma^2 \left( d\tau^2 + a(\tau)^2 d\Omega_{3}^2 \right)
\ea
where is $\sigma^2 = {2G}/{3 \pi}$. With this metric ansatz, the 
action becomes
\ba
S_{E} = \frac{1}{2}\int d\tau \left( -a\dot{a}^2 - a + \lambda a^3 \right)
\ea
where $\lambda = {\sigma^2 \Lambda}/{3} = {2G \Lambda}/{9 \pi}$. So
$\lambda$, $\tau$, and $a$ are all dimensionless.

%\section{The deSitter Spacetime and its Perturbations}

\subsection{The deSitter Space}

In this section we give a summary of the result of \cite{Halliwell:1985eu}.
Consider a compact three-surface $S^3$ which divides the four-manifold M into
two parts. One can introduce the coordinates $x^i$ ($i = 1,2,3$) and a coordinate $t$ 
such that $S^3$ is the surface at $t = 0$. The metric takes the form
\ba
ds^2 = -(N^2 - N_i N^i)dt^2 + 2N_idx^idt + h_{ij}dx^idx^j.
\ea
where $N$ and $N_i$ are the lapse function and the shift vector, respectively.
The action is given by
\ba
S = \int \left( L_g + L_m \right) d^4x
\ea
where
\ba
L_g = \frac{M_{P}^{2}}{16\pi}N \left( G^{ijkl}K_{ij}K_{kl}
+ \sqrt{h} R  \right)
\ea
where $R$ is the Ricci scalar of the three-surface and $K_{ij}$ is
the second fundamental form given by
\ba
K_{ij} = \frac{1}{2N}\left( -\frac{\partial h_{ij}}{\partial t}
+ 2N_{(i|j)}  \right)
\ea
In the above expression ``$|$'' denotes the covariant derivative.
$G^{ijkl}$ is called the metric of the ``superspace'' and is given by
\ba
G^{ijkl} = \frac{1}{2}\sqrt{h}\left(h^{ik}h^{jl}+ h^{il}h^{jk}
- 2h^{ij}h^{kl} \right)
\ea
In the case of a massive scalar field, the matter Lagrangian $L_m$
is given by
\ba
L_m = \frac{1}{2}N\sqrt{h}\left( N^{-2}\left(\frac{\partial \Phi}
{\partial t} \right)^2 -2\frac{N^i}{N^2}\frac{\partial \Phi}
{\partial t} \frac{\partial \Phi}{\partial x^i} 
\right . \\ \nonumber \left. - \left( h^{ij} 
-\frac{N^i N^j}{N^2}\right)\frac{\partial \Phi}
{\partial x^i}\frac{\partial \Phi}{\partial x^j} - m^2 
\Phi^2  \right)
\ea
In the Hamiltonian treatment of general relativity one treats $h_{ij}$
and $\Phi$ as the canonical coordinates. The canonically conjugate
momenta are
\ba
\pi^{ij} = \frac{\partial L_g}{\partial\dot{h_{ij}}}
= -\frac{\sqrt{h}M_P^2}{16\pi}\left(K^{ij} - h^{ij} K \right)\\
\nonumber
\pi_{\Phi} = \frac{\partial L_m}{\partial \dot{\Phi}}
= N^{-1}\sqrt{h}\left(\dot{\Phi}-N^i \frac{\partial \Phi}{\partial x^i}   
\right)
\ea
The Hamiltonian is
\ba
H = \int d^3x \left(\pi^{ij}\dot{h_{ij}} + \pi_{\Phi}
\dot{\Phi}-L_g -L_m \right) \\ \nonumber
= \int d^3x \left(NH_0 + N_i H^i \right)
\ea
where
\ba
H_0 = 16 \pi M_P^{-2}G_{ijkl}\pi^{ij}\pi^{kl}  - \frac{M_P^{2}}{16\pi}
\sqrt{h}R + & \\ \nonumber\frac{1}{2}\sqrt{h}\left(\frac{\pi_{\Phi}^2}{h}
+ h^{ij} \frac{\partial \Phi}{\partial x^i}
\frac{\partial \Phi}{\partial x^j} 
 + m^2 \Phi^2 \right) 
\ea
\ba
H^i = -2\pi^{ij}_{|j} + h^{ij}\frac{\partial \Phi}{\partial x^j}
\pi_{\Phi} &&
\ea
and
\ba
G_{ijkl} = \frac{1}{2}h^{-1/2}\left(h_{ik}h_{jl}+h_{il}h_{jk}-
h_{ij}h_{kl} \right)
\ea
The quantities $N$ and $N_i$ are regarded as Lagrange multipliers.
Thus the solution obeys the momentum constraint
\ba
H^i = 0
\ea
and the Hamiltonian constraint
\ba
H_{0} = 0.
\ea

\subsection{The Perturbations}

Now we study the perturbations around the deSitter spacetime. The perturbed
deSitter has a three-metric $h_{ij}$ of the form
\ba
h_{ij} = a^2 \left( \Omega_{ij}+\epsilon_{ij}\right)
\ea
where $\Omega_{ij}$ is the metric on the unit three-sphere and 
$\epsilon_{ij}$ is a perturbation on this metric and can be expanded
in harmonics:
\ba
\label{harmo}
\epsilon_{ij} = \sum_{n,l,m} \left[ \sqrt{6}a_{nlm}\frac{1}{3}
\Omega_{ij} Q^n_{lm} + \sqrt{6}b_{nlm}(P_{ij})^n_{lm} +\right . &  \nonumber \\
\left .
\sqrt{2}c^0_{nlm}(S^0_{ij})^n_{lm} + \sqrt{2}c^e_{nlm}(S^e_{ij})
^n_{lm} + 2t^0_{nlm}(G^0_{ij})^n_{lm}
+ 2t^e_{nlm}(G^e_{ij})^n_{lm}
\right]
\ea
where the coefficients $a_{nlm}$, $b_{nlm}$, $c^0_{nlm}$, $c^e_{nlm}$,
$t^0_{nlm}$ and $t^e_{nlm}$ are functions of time but not the three 
space coordinates. 
The $Q(x^i)$ are the scalar harmonics on the three-sphere. The
$P_{ij}(x^i)$ are given by
\ba
P_{ij} = \frac{1}{n^2 -1}Q_{|ij}+ \frac{1}{3}\Omega_{ij}Q
\ea
where we have suppressed the $n,l,m$ indices. The $S_{ij}$ are
given by
\ba
S_{ij} = S_{i|j} + S_{j|i}
\ea
where $S_{i}$ are the transverse vector harmonics. The $G_{ij}$
are the transverse traceless tensor harmonics. Further details 
can be found in \cite{Halliwell:1985eu,Gerlach:1978gy}.
The lapse, shift, and the scalar field $\Phi(x^i,t)$ can be
expanded in terms of harmonics as
\ba
N = N_0 \left[1 + \frac{1}{\sqrt{6}}\sum_{n,l,m}g_{nlm}Q^n_{lm} \right]
\\ \nonumber
N_i = a(t)\sum_{n,l,m}\left[\frac{1}{\sqrt{6}}k_{nlm}(P_i)^n_{lm}
+ \sqrt{2}j_{nlm}(S_i)^n_{lm} \right] \\ \nonumber
\Phi = \sigma^{-1}\left[\frac{1}{\sqrt{2}\pi}\phi(t)+ \sum_{n,l,m}f_{nlm}Q^n_{lm} \right]
\ea
where $P_i = \frac{1}{n^2 -1}Q_{|i}$. The perturbed action is
now given by
\ba
\label{act}
S = S_{0}(a, \phi, N_0)+ \sum_{n} S_n
\ea
 where we have denoted the labels $n$, $l$, $m$, $o$, and $e$ by the single
label $n$. $S_0$ is the action of the unperturbed deSitter,
\ba
S_{0} = -\frac{1}{2}\int dt N_0 a^3\left(\frac{\dot{a}^2}{N_o^2a^2} 
- \frac{1}{a^2}- \frac{\dot{\phi}^2}{N_0^2}+ m^2\phi^2 + \lambda
 \right) 
\ea
$S_n$ is quadratic in perturbations and is given by
\ba
S_n = \int dt (L^n_g + L^n_m)
\ea
where $L^n_g$  and $L^n_m$ are the $n$th mode gravitational and
the matter Lagrangians, respectively. As we are interested only
in the gravitational tensor and the scalar field perturbations
in this paper, we only display those terms here. In the absence of sources, the 
other modes can be gauged away. For further
details and the intricacies of all perturbations we refer the
reader to \cite{Halliwell:1985eu}. 

The tensor perturbations $t_n$ have the Euclidean action
\ba
S_n^E = \frac{1}{2}\int t_n \hat{D} t_n +  ~boundary ~term
\ea
where
\ba
\hat{D} = \left(- \frac{d}{d\tau}\left[a^3\frac{d}{d\tau} \right]
+ a(n^2 - 1)  \right)
\ea
where the background satisfied the classical equation of motion.
The action is extremized when $t_n$ satisfies the equation
$\hat{D} t_n = 0$. Setting $N_0=1$, this is just the equation
\ba
\label{eom_ten}
\frac{d}{dt}\left[a^3 {\dot{t_n}} + (n^2 -1)a t_n\right] = 0
\ea
For $t_n$ that satisfies the equation of motion, the action
is just the boundary term
\ba
\label{E_cl}
S_n^{E(cl)} = \frac{1}{2}a^3 \left(t_n\dot{t_n}+4\frac{\dot{a}}{a}
t_n^2  \right)
\ea 
The path integral over $t_n$ will be
\ba
\label{tensor}
\int d[t_n]\exp (-S_n^{E(cl)})= (\det \hat{D})^{-1/2}\exp (-S_n^{E(cl)})
\ea

Scalar fields can be treated in a similar fashion. The scalar field perturbation Lagrangian is given by
\ba
\frac{1}{2}N_0 a^3 \left[\frac{1}{N_0^2}\dot{f_n}^2 -m^2 f_n^2 -\frac{(n^2 -1)}{a^2}f_n^2 \right]
\ea
where we work in an appropriate gauge choice. 
%Also we have assumed that the scalar field kinetic term is negligible. 
Setting $N_0=1$, the equation of motion for $f_n$ is
\ba
 \frac{d}{dt}\left(a^3 {\dot{f_n}}\right) + \left[m^2 a^3 + (n^2 -1 )a \right]f_n = 0
\ea
One can evaluate the scalar field path integral and the results are 
similar to that of the tensor perturbative modes. So in this paper
we deal only with the gravitational tensor modes.

\section{Perturbative Correction to the Bounce : No Back Reaction}

Before doing the calculation that includes the backreaction due to the
perturbative modes, we first perform a simple calculation to illustrate the key issues we are facing.
The result of this section is meant to be a warm up and indicates the possibility
of a correction to the wavefunction. We consider a $S^4$ solution with both its polar regions flattened by a small parameter $\delta$ and show that this squashed geometry allows for extra perturbative modes that can have significant effect on the calculation of the wavefunction. When $\delta = 0$, these extra perturbative modes vanish and what remains is the well known one-loop correction to the $S^4$.  For a proper treatment the reader
may go directly to the next section and onward.\\

 The Euclidean equation of motion for the $n$th tensor perturbation mode
on a $S^4$ background follows from Eq.(\ref{eom_ten})
\ba
\label{t_n}
\ddot{t_n}+ 3\frac{\dot{a_o}}{a_o}\dot{t_n} - 
\left( \frac{(n^2 -1)}{a_o^2} \right)t_n  = 0
\ea
where dot denotes differentiation with respect to the $\tau$ variable and
$a_o(\tau) = \lambda^{-1/2} \sin(\sqrt{\lambda}\tau)$.
This can be converted to a more familiar form by the substitution
$F_n = \sqrt{\lambda} a_o(\tau) t_n$.
In terms of $F_n$ and $x = \cos(\lambda \tau)$ the equation of motion becomes
\ba
(1 - x^2)\frac{d^2 F_n}{dx^2} - 2x\frac{dF_n}{dx} +
\left( 2 - \frac{n^2}{(1 - x^2)} \right) F_n = 0 
\ea
which is just an associate Legendre equation of degree one
and order $n$. This has two linearly
independent solutions, $P^{-n}_{1}(x)$ and $Q^{n}_{1}(x)$.
However, because $a_o=0$ at $x=\pm 1$, we have
\ba
S_{E}^{n} & = & \frac{1}{2} \left[a(\tau)^3 t_n \frac{dt_n}{d\tau}  
\right]_{x = 1}^{x = -1} =0
\ea

We shall see in the following sections that once the perturbative modes
are properly accounted for, $S^4$ changes to a ``barrel''. Anticipating
this result we consider the $S^4$ instanton that is flattened slightly
at the two poles. This ``squashed'' $S^4$
is given by $a_o(\tau) = \frac{1}{\sqrt{\lambda}}\sin(\lambda \tau)$,
with the restricted range $ 1 - \delta \geq x \geq -1 + \delta$. There
is a good reason for considering the squashed geometry. The perturbative
modes can be strong enough to change the geometry of $S^4$ (and as we shall
see, they will). So fixing the geometry to $S^4$ before doing the
perturbative analysis is too restrictive. The perturbative modes 
on $S^4$  have to vanish at the two poles if they are to respect the
background geometry. Their effect has been studied in \cite{Barvinsky:1992dz
,Gibbons:1978ji} and apart from contributing to one-loop effect 
they do not lead to any tunneling suppression.
 However, to get the decoherence effect  one must
include modes that have nonvanishing values at the two poles. And
we shall see that the squashed geometry allows for modes that
can potentially lead to decoherence. This expectation will be confirmed in
 the later sections. \\

     As explained in the introduction (Eq.(\ref{path},\ref{trace})), to 
trace out a given  perturbative mode
(say, $t_n$) one must perform a path integral over that mode with
the initial and final amplitudes the same (say, $t_n^i = t_n^f$), and
then integrate over all possible values of $t_n^i$. This is just
the trace operation in the path integral formalism. To do this we
first find the action for $t_n$ (details can be found in the appendix)

\ba
S_{E}^{n} & = & \frac{1}{2} \left[a(\tau)^3 t_n \frac{dt_n}{d\tau}  
\right]_{x = (1 - \delta)}^{x = (-1 + \delta)} \\ \nonumber
& = & \delta(2 -\delta) \left[ (1 - \delta) + \delta(2 -\delta)
\frac{\partial_{x}Q^{n}_1(1 -\delta)}{Q^{n}_1(1 -\delta)}\right] (t_n^i)^2
\ea
Next we must find the prefactor to the path integral for $t_n$.
This can be found as explained in the Appendix B. The prefactor 
is given by
\ba
\frac{1}{\sqrt{4\pi}} \frac{\sqrt{Q^n_1(1 -\delta) P^n_1(1 - \delta) 
(2\delta- \delta^2)}}{2^n} \sqrt{\frac{\Gamma\left( \frac{2-n}{2}
\right) \Gamma\left( \frac{3-n}{2}\right)}{\Gamma\left( 
\frac{2+ n}{2}\right) \Gamma\left( \frac{3+n}{2}\right)} }
\ea
Integrating over all initial states $t_n^i$ (tracing out the $nth$ mode)
gives
\ba
\int dt_n^i K \left( t_n^i, x = -1 + \delta ;t_n^i, x = 1 -\delta \right)  
= \alpha (n) f(\delta) 2^{-n}
\ea
where $\alpha(n)$ contains factors of $n$ and $f(\delta)$ is a function 
of $\delta$, with $f(0)=0$. $K \left( t_n^i,-1 + \delta ;t_n^i,1 -
\delta \right)$ is the path integral over $t_n$ mode with the boundary 
conditions
$t_n(x = -1 + \delta) = t_n(x = 1 - \delta) = t_n^i$.
Taking care of all the modes by tracing over all of them (within the
lower and upper cut-offs), we get
\ba
\prod_{n}\int dt_n^i  K \left( t_n^i, -1 + \delta ;t_n^i, 1 -\delta \right)
= \prod_{n}
\alpha (n) f(\delta) 2^{-n} \\ \nonumber
= A(n, \delta)e^{-N^4/2 \ln(2)}
\ea
where $N$ counts the modes. As explained later,
%in Eq.(\ref{N}, \ref{nu})
\ba
N = \left( \frac{H^{-1}}{l_s}\right) = 2^{1/4}\frac{\nu^{1/4}}{\sqrt{\lambda}}
\ea
where $\nu = \frac{2}{9\pi^2}\frac{G^2}{l_s^4}$. The 
wavefunction then corresponds to the path integral over the scale factor
$a$ and the perturbative modes $t_n$ with the full action as given
in Eq.(\ref{act}). We trace over the perturbative modes and get
\ba
\Psi \simeq e^{\left(\frac{1}{3\lambda} - D \right)}
\ea
where $D$ is the decoherence term leading to tunneling suppression
\ba
D \simeq \frac{\nu}{\lambda^2}
\ea
so the above result leads 
to an order O(${1}/{\lambda^2})$ suppression to the Hartle-Hawking
wavefunction. This naive (naive because, as it turns out, the backreaction
is important and also leads to an O(${1}/{\lambda^2})$ contribution)
expectation is indeed vindicated in our calculation of the modified bounce.
The more careful calculation changes the coefficient $\ln(2)$, but maintains
the inverse square dependence on $\lambda$.

\section{The Modified Bounce : Including Backreaction}

The bounce solution $S^4$ is a solution to the Euclidean Einstein
equation which is obtained as the Euler-Lagrange equation from
the Euclidean action $S^E_{0}[a]$. 
\ba
\label{euc_eins}
-2\frac{\ddot{a}}{a} - \frac{\dot{a}^2}{a^2}+\frac{1}{a^2}
= 3 \lambda
\ea
The solution is given by $a(\tau) = 
\sqrt{\frac{1}{\lambda}}\cos(\sqrt{\lambda}\tau)$. This is 
the bounce in the absence of any perturbation. Including
the perturbations, treating them as the environment, and tracing
them out (as explained in Eq.(\ref{path},\ref{trace})) will lead to 
a modified equation of motion instead
of Eq.(\ref{euc_eins}). In this section we derive this modified
bounce equation.

Let us consider the effect of the metric perturbations. To find 
the modified bounce equation we have to carry out the
path integral over the perturbation modes $t_{n}$ is Eq.
(\ref{tensor}). Let us write down the path integral for a single
tensor perturbation mode
\ba
\label{PI}
\int D[t_{n}(\tau)]\exp \left( -\frac{1}{2}\int_{-T/2}^{T/2}
d\tau a^3 [ \dot{t_{n}}^2 + \frac{(n^2 - 1)}{a^2} 
t_{n}^2 ] \right)
\ea
This is a path integral for an oscillator with a varying
mass as well as frequency. We can simplify this to a path
integral of an oscillator with constant mass and variable
frequency (given by $\omega_n (a(u)) = \sqrt{(n^2 -1)}a(u)^2$) 
using a new variable $u$
\ba
du = \frac{d\tau}{a(\tau)^3}
\ea
The Euclidean action now becomes
\ba
\label{action}
S_{E} = \frac{1}{2}\int_{u_i}^{u_f}du \left( t_{n}'^2 +
 (n^2 - 1)a^4t_{n}^2\right) \\ \nonumber
= \frac{1}{2}\int_{u_i}^{u_f} du \left( -tt'' + \omega_n^2 t^2
\right) + \frac{1}{2} (t{t}'){\Large\bf{ |}}^{u_f}_{u_i}
%\int_{u_i}^{u_f} du \frac{d}{du}\left(tt'\right)
\ea
To keep notation uncluttered, we drop the subscript $n$ for
the time being. Let $t = t_{cl} + \hat{t}$, where
$t_{cl}$ is a solution to the classical equation of motion
for the above action
\ba
\label{eom}
t_{cl}'' - \omega_{n}(a)^2 t_{cl} = 0
\ea
with $t_{cl}(u_i)=t(u_i)$ and $t_{cl}(u_f)=t(u_f)$. 
Here $\hat{t}$ denotes fluctuations about
the classical solution with $\hat{t}(u_i)=\hat{t}(u_f)=0$. That is,
$(t {t}'){\Large |}^{u_f}_{u_i}= (t_{cl}\hat{t}'){\Large |}^{u_f}_{u_i} + (t_{cl}{t_{cl}}'){\Large |}^{u_f}_{u_i}$.
Apriori, the contributions to the path integral 
will come from $t_{cl}$ and $\hat t$.
Here, the fluctuations $\hat{t}$ will lead to a prefactor. We shall keep
track of this prefactor as it will have important contribution
to the modified action. Substituting $t = t_{cl} + \hat{t}$ in 
 Eq.(\ref{action}), we obtain
\ba
\label{actionI}
S_E = \frac{1}{2}\int_{u_i}^{u_f} du \left( -\hat{t}\hat{t}''
+ \omega_n^2 \hat{t}^2 \right) + (t_{cl}{t_{cl}}'){\Large\bf{ |}}^{u_f}_{u_i}
%\frac{1}{2}\int_{u_i}^{u_f} du \frac{d}{du}(t_{cl}t_{cl}')
\ea
The second term in the above equation is simply $S_E(t_{cl})$.
The path integral in Eq.(\ref{PI}) is, therefore, given by
\ba
\label{PII}
e^{-S_E(t_{cl})} P[{\hat{t}}]=e^{-S_E(t_{cl})} \int^0_0 D[\hat{t}(u)] \exp \left( -\frac{1}{2}\int_{u_i}^{u_f} 
du \left( -\hat{t}\hat{t}''+ \omega_n^2 \hat{t}^2 \right)\right)
\ea
Notice that the prefactor $P[{\hat{t}}]$ is independent of $t(u_i)$ or $t(u_f)$.
So the reduced path integral Eq.(\ref{trace}) becomes
\ba
\label{trace2}
Tr[Z] =  \int D[a] \prod_{n} P[{\hat{t_n}}] \int dt_n^i \int dt_n^f \delta(t_n^i - t_n^f) e^{-S_E(t_{n,cl})}
\ea
We shall first evaluate the integral over $t_n^i=t_n^f$. In Appenxix C we evaluate the prefactor 
$P[{\hat{t_n}}]$. It is straightforward to include a scalar field.

Although finding a general solution to Eq.(\ref{eom}) is
in general impossible, we note that it has the
same form as the Schrodinger equation for a particle in
a potential $\omega_n (a(u))$. So for a slowly varying
potential we can use the WKB method for finding the solution
to Eq.(\ref{eom}). By slowly varying one means that the
following condition is satisfied
\ba
\frac{d\omega_{n}}{d\tau} << \omega_{n}^2
\ea
This is satisfed by the higher modes $n >> 1$ and we 
shall see that these modes are the ones that contribute  
to the suppression of quantum tunneling. 
(One may be concerned that $ \omega_{n} \to 0$ as $a \to 0$.
As we shall see, in contrast to the unperturbed bounce, $a$ actually stays 
finite in the modified bounce. In this sense, the WKB approximation is 
reasonable.)
The two independent WKB solutions to Eq.(\ref{eom}) are
\ba
\label{class_soln}
t_{cl}^{\pm} = \frac{1}{\sqrt{\omega_n}}\exp \left( \pm 
\int^u du' \omega_n(u')\right)
\ea
The general solution will be a linear combination of the
independent solutions
\ba
\label{gensol}
t_{cl}(u) = At_{cl}^{+}(u) + Bt_{cl}^{-}(u)
\ea
The values of $A$ and $B$ will depend on the boundary
conditions. If Eq.(\ref{gensol}) is to satisfy the 
following boundary conditions
\ba
t_{cl}(u_i) = t^i &  t_{cl}(u_f) = t^f
\ea
then we have the following values of $A$ and $B$
\ba
\label{AB}
A = \frac{t^f \sqrt{\omega_n(u_f)} - t^i 
\sqrt{\omega_n(u_i)}\exp(- D_n)}{(\exp(D_n) - \exp(-D_n))} 
\\ \nonumber
B = -\frac{t^f \sqrt{\omega_n(u_f)} - t^i 
\sqrt{\omega_n(u_i)}\exp(D_n)}{(\exp(D_n) - \exp(-D_n))}
\ea
where $D_{n}$ is given by
\ba
\label{D_neq}
D_n = \int_{u_i}^{u_f}du ~ \omega_{n}(a(u)) = 
\int_{-T/2}^{T/2}d\tau~ \frac{\omega_{n}(a(\tau))}{a(\tau)^3}
\ea
Next, substituting the solution Eq.(\ref{gensol}), using 
Eq.(\ref{AB}), in Eq.(\ref{actionI}), we get
\ba
\label{S_E}
& S_{E}(t_{cl}) = \frac{1}{2} \left( t_{cl}(u_f)\frac{dt_{cl}(u_f)}{du} - 
t_{cl}(u_i)\frac{dt_{cl}(u_i)}{du} \right)   \\ \nonumber
& = \frac{1}{2(\exp(D_n) - \exp(-D_n))} \left [\left( (t^f)^2 
\omega_n(u_f) +(t^i)^2 \omega_n(u_i) \right) \left( \exp(D_n) 
+ \exp(-D_n)\right) \right. \\ \nonumber &  \left. - 4t^i t^f 
\sqrt{\omega_n(u_i)\omega_n(u_f)}
\right]
\ea 
This gives the classical contribution Eq.(\ref{PII}). Note that
in order to perform the trace over this perturbation mode, we
will be setting $t^i = t^f$, and $\omega_n(u_i) = \omega_n(u_f)$.
The path integral over $\hat{t}$ leads to the prefactor. Its evaluation
for a time-dependent frequency is explained in Appendix C (Eq.(\ref{go})).  
(Equivalently, one may use the Pauli-van Vleck-Morette formula  
studied by Barvinsky \cite{Barvinsky:1993nf}). 
The prefactor (path integral over $\hat{t}$) gives
\ba
\label{pre_fac}
\int D[\hat{t}(u)] \exp \left( \frac{1}{2}\int_{u_i}^{u_f} 
du \left( -\hat{t}\hat{t}'' + \omega_n^2 \hat{t}^2 \right)\right)
=  \frac{\sqrt{\omega_n(u_f)}}{\sqrt{\left(\exp(D_n) - \exp(-D_n)\right)}}
%\frac{1}{\sqrt{(\exp(D_n) - \exp(-D_n))}}
\ea
Now the trace over the perturbative modes can be performed by
setting $t^i = t^f$ and integrating over the amplitudes $t^i$
\ba
& \int dt^f \int dt^i \delta(t^i -t^f)\int_{t^i}^{t^f}D[t(u)] \exp 
\left( -S_{E}\right) \\ \nonumber
= &  \frac{\sqrt{\omega_n(u_f)}}{\sqrt{(\exp(D_n) - \exp(-D_n))}} 
\int dt^f \exp\left( -S_{E} [t_{cl}] \right)
\\ \nonumber
\simeq  & \frac{1}{\sqrt{\left(\exp(D_n) - \exp(-D_n)\right)}}
\ea
where we used $\int dt^f \exp\left( -S_{E} [t_{cl}]\right) \simeq 1/\sqrt{
\omega_n(u_f)}$. Also, note that $D_n >> 1$ for large values of $n$, and 
$\left(\exp(D_n) -  \exp(-D_n)\right)^{-1/2} \simeq \exp (-D_n/2)$ . 
Finally, one has to  perform this tracing-over over all $n$ modes. 
This leads to the following contribution
\ba
\prod_{n} \frac{1}{\sqrt{(\exp(D_n) - \exp(-D_n))}}\simeq \exp (- \sum\frac{D_n}{2} )
\ea
 From the expansion in Eq.(\ref{harmo})
it is clear that one has to count the indices $n, l, m$  (corresponding 
to the spherical harmonics on $S^3$). For a given $n$, $l$ can take 
$n-2$ values ($l = 2, 3, ..., n-1$), and for a given $l$, $m$ can take
$2l + 1$ values ($m = -l, -l + 1, ...., l - 1, l$). For a given $n$ this
introduces a degeneracy of $f(n) \simeq 2 n^2$. The factor of $2$ comes \
due to the two tensor modes
$t^0_{nlm}$ and $t^e_{nlm}$ in the expansion in Eq.(\ref{harmo}).
To proceed further, we need to define the cut-offs. A natural short wavelength 
cut-off is the string scale, $l_{s}$. The long wavelength cut-off
is the inverse Hubble length.
%The number of modes between these two cutoffs is given by $N$
This gives a cut-off for $n$
\ba
\label{N}
n_{max} = N = \left( \frac{H^{-1}}{l_s}\right)=\frac{1}{l_s}\sqrt{\frac{3}{\Lambda}}
\ea
\ba
D \equiv \sum \frac{D_n}{2}  = \frac{1}{2} \sum^N_{n,l,m} \int_{-T/2}^{T/2}
d\tau \frac{\omega_n(a)}{a^3} 
\ea 
where we have used Eq.(\ref{D_neq}).
Note that 
$\omega_n(a)^2 = (n^2 - 1)a^4$. For higher modes, therefore,
$\omega_n \simeq na^2$. Thus,
\ba
D = \frac{1}{2}\int_{-T/2}^{T/2}\frac{d\tau}{a} \sum^N_{n,l,m} n \\ \nonumber
= \frac{N^4}{8}\int_{-T/2}^{T/2}\frac{d\tau}{a}=\frac{1}{2}\int_{-T/2}^{T/2}{d\tau}\frac{\nu}{\lambda^2a}
\ea
We can also easily generalize to arbitrary number of degrees of freedom, with
\ba
\frac{\nu}{\lambda^2}= \frac{n_{dof} N^4}{4} = \frac{n_{dof}}{4(Hl_s)^4}
\ea
For pure gravity, we have $n_{dof}=2$.
The modifed Euclidean action is, therefore, given by
\ba
\label{bounce}
S_{E,dC}= S_{E,0} + D[a] = \frac{1}{2}\int d\tau
\left( -a\dot{a}^2 - a + \lambda a^3 + \frac{\nu}{\lambda^2 a} \right)
\ea
To evaluate $S_{E,dC}$ via the saddle-point approximation, we have to find the classical path, or the bounce solution, of this effective action.

\section{The Bounce Solution}

The equation of motion for the Euclidean action,
Eq.(\ref{bounce}), is
\ba
\label{modbounce}
-2\frac{\ddot{a}}{a} - \frac{\dot{a}^2}{a^2}+\frac{1}{a^2}
= 3\lambda - \frac{\nu}{\lambda^2 a^4} 
\ea
This is just the Euclidean version of the Einstein equation with
both a cosmological constant and radiation. As is well known, this 
equation allows for a variety of solutions (M1, M2, A1, A2, E, O1)
\cite{Harrison:1967ab}. One has to be careful in deciding which one
of these is the correct bounce solution.The bounce solution is the 
real solution to the classical Euclidean equation of motion. The M2 
solution satisfies the bounce criteria and is given by 
(in the Euclidean form)
\ba
\label{bouncesol}
a(\tau) = \frac{1}{\sqrt{2\lambda}} \sqrt{\left( 1 
+ \sqrt{\left( 1 - \frac{4 \nu}{\lambda}\right)} \cos 
(2 \sqrt{\lambda}\tau) \right)}
\ea
This solution is the modified bounce and has some interesting 
features.

\begin{figure}
\begin{center}
\epsfig{file=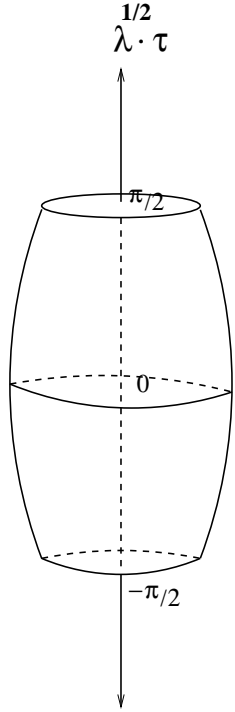, width=3cm}
\vspace{0.1in}
\caption{The Modified Bounce : It is a squashed version of $S^4$ resulting in a ``barrel''.}
\label{fig6}
\end{center}
\end{figure}

\begin{itemize}
\item
It is a deformation of the $S^4$ instanton from spherical to barrel-shaped,
with $ - \pi/2 \sqrt{\lambda} \le \tau \le  \pi/2 \sqrt{\lambda}$.
It reduces to the usual $S^4$ instanton when $\nu = 0$, as expected
\ba
a(\tau) \to \frac{1}{\sqrt{\lambda}} \cos (\sqrt{\lambda} \tau)
\ea
Here $\nu$ is the deformation parameter.
\item
For large values of $\nu$ the bounce is destroyed. Presumably, there 
is no more
quantum tunneling because of excessive decoherence. This critical
value of $\lambda$ is given by $\lambda_c= 4 \nu $. 
For tunneling $\lambda > \lambda_c$.

%That the deformed $S^4$ has to do with the lifting of ``nothing''  
%can also be seen by noting that the modified bounce corresponds
%to a modified Wheeler-DeWitt equation or, what is equivalent, a
%modified Hamiltonian constraint. This, and the consequence this has
%for the boundedness of the gravitational potential, we explain in the next section. 

\end{itemize}

\begin{figure}
\begin{center}
\epsfig{file=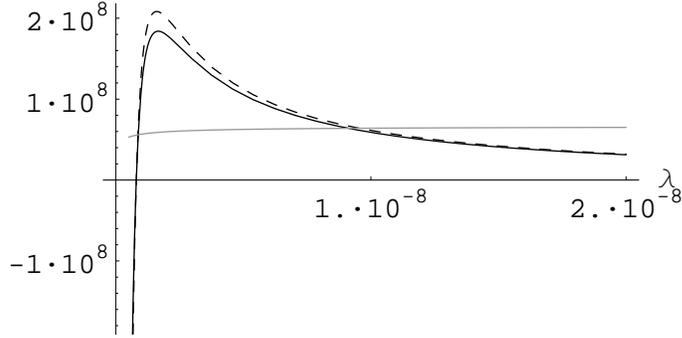, width=9cm}
\vspace{0.1in}
\caption{The solid curve is $F= - S_{E,dC}$, 
using the exact value of $\Im$. The dotted curve shows $F$ using 
$\Im \simeq 8/3$. The light flat curve is the 
integral $\Im$, which asymptotes to $8/3$ for large $\lambda$. 
The plots are given with $M_{Pl}/M_s = 10^3$.
The dotted and the solid curves differ only slightly, 
justifying the approximation made.}
%{The solid curve is the negative modified action, $-F$, 
%(Eq.(\ref{modac})) using the exact value of $\Im$ (Eq.(\ref{im1})). 
%The dotted curve shows $-F$ using approximation in Eq.(\ref{im2}). 
%The light flat curve shows the shape of the integral $\Im$. This 
%justifies the approximation made.}
\label{fig7}
\end{center}
\end{figure}

The modified Euclidean action can now be calculated. One has to be
careful though to include the O$({\nu}/{\lambda^2})$ contribution
from the $-a\dot{a}^2 - a + \lambda a^3$ part. This contribution
 will add to the O$({\nu}/{\lambda^2})$ contribution from the
$\frac{\nu}{\lambda^2 a(\tau)}$ term and give the total O$({\nu}/{\lambda^2})$ 
modification. We state the final result
\ba
\label{modac}
-F =S_{E,dC} =\left( -\frac{1}{4\lambda} + 
\frac{\nu}{\lambda^2}\right) \Im
\ea 
where $\Im$ is an integral given by
\ba
\label{im1}
\Im = \int_{\frac{-\pi}{2\sqrt{\lambda}}}^{\frac{\pi}
{2\sqrt{\lambda}}}d\tau \left( \frac{\sin^2(2\sqrt
{\lambda}\tau)}{a(\tau)} \right)
\ea
 where $a(\tau)$ is the bounce solution given by Eq.(\ref{bouncesol}). 
 The exact value of $\Im$ involves
an elliptic integral, but we can make a fairly accurate 
estimate and the result is
\ba
\label{im2}
\Im \simeq \frac{8}{3}
\ea

For tunneling from nothing, we should consider the barrel to be a deformed $S^4$ with 
the same topology. That is, we should include the contribution of end plates of the barrel 
in the evaluation of $S_{E,dC}$ in Eq.(\ref{modac}). At the end plates, 
$$ a = \frac{1}{\sqrt{2 \lambda}} \left(1-\sqrt{1 - \frac{4\nu}{\lambda}}\right)^{1/2} > 0$$
However, their contribution to $S_{E,dC}$ is zero because the weighing factor $\sin^2(2\sqrt
{\lambda}\tau)$ in Eq.(\ref{im1}) vanishes at the end plates.

As expected, for $\nu = 0$ we get the usual Hartle-Hawking
wavefunction. Now, including the environment,
%Up to a O$({\nu}/{\lambda^2})$, however, the 
the tunneling probability is given by 
%modified wavefunction is just given by
\ba
\label{modwf}
P \simeq e ^{\left(\frac{2}{3\lambda} - \frac{8\nu}{3\lambda^2}\right)}
\ea
For $\lambda \to 0$, higher order decoherence effects should be included.
These effects can come from a more careful treatment of the leading term
we have obtained, or from higher order interaction of the modes with $a$
and among themselves. Quantum effects may become relevant. This is 
especially so at large $\lambda$. Note that $S_{E,dC}$ peaks around
$\lambda = 8 \nu$. So for regions with non-zero tunneling probability, 
$\lambda > \lambda_{c} = 4\nu$.

\section{The Modified Hamilton Constraint and the Wheeler-DeWitt Equation}

The modified action, as given in Eq.(\ref{bounce}), leads to a modified
Hamiltonian constraint. The modified Lorentzian action is given by
\ba
S = \frac{1}{2}\int d\tau \left( - a\dot{a}^2 + a - \lambda a^3 
- \frac{\nu}{\lambda^2 a} \right)
\ea 
The modified Hamiltonian constraint is
\ba
H = \frac{1}{2a}\left( -\Pi_{a}^2 - a^2 + \lambda a^4  
+ \frac{\nu}{\lambda^2} \right) = 0
\ea
where $\Pi_a = -a\dot{a}$ is the conjugate momentum. 
Using the Hamiltonian constraint and the equation of motion following from the modified 
Lorentzian action, we have 
\ba
(\frac{\dot a}{a})^2 + \frac{1}{a^2} = \lambda + \frac{\nu}{\lambda^2 a^4} \\ \nonumber
\frac{\ddot a}{a} = \lambda - \frac{\nu}{\lambda^2 a^4}
\ea
Since ${\ddot a}/a= - \sum (\rho_i +3p_i)/2=-\sum \rho_i (1 +3\omega_i)/2$, this implies that 
the new $\nu$ term has equation of state $\omega=1/3$, precisely that of ordinary radiation. 

\begin{figure}
\begin{center}
\epsfig{file=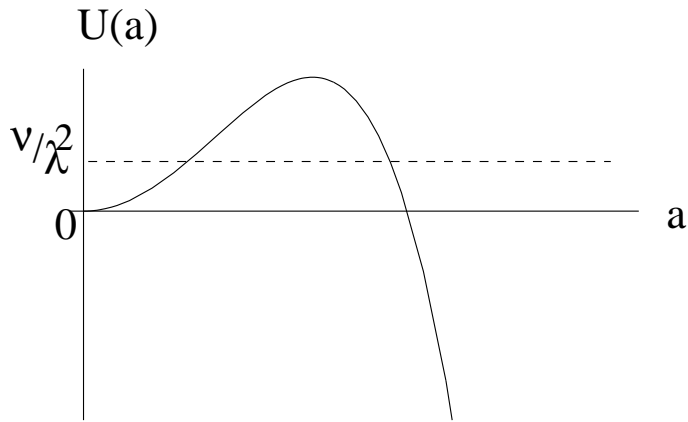, width=9cm}
\vspace{0.1in}
\caption{The solid curve is $U_{0}(a)=a^2 -\lambda a^{4}$.
The effect of the new term is simply to raise the energy eigenvalue from $0$ to $E=\nu/\lambda^2>0$.} 
\label{fig8}
\end{center}
\end{figure}

One can write down a differential equation describing the evolution
of the wavefunction - the Wheeler-DeWitt (WdW) equation -by imposing 
the condition as an operator equation. This is quantum gravitational analog of
the Schrodinger equation in quantum mechanics.
\ba
\label{wdw}
-\hat{H}\Psi =
% 0 \nonumber \\
\left( \frac{\hat{\Pi_{a}}^2}{2} + U(a)\right)\Psi = 0 \nonumber 
\ea
where $\hat{\Pi_a} = -i\partial / \partial a$ and $U(a)$ is 
the gravitational potential given by
\ba
U(a) = U_0(a)  - \frac{\nu}{\lambda^2} = a^{2} - \lambda a^4  - \frac{\nu}{\lambda^2}
\ea
Recall the case without the $\nu$ term. 
In the classically forbidden region, i.e., the under-barrier region ${\lambda}^{-1/2} \ge a \ge 0$, 
the WKB solutions for the tunneling amplitude from $a=0$ to $a=\lambda^{-1/2}$ are
\ba
\label{HKWKB}
 \Psi_{\pm} \simeq e^{\pm \int_0^{1/\sqrt{\lambda}} |\hat \Pi(a')| da'}
\ea
%Because the kinetic term in Hamiltonian $\hat H$ is negative. 
Note the Hartle-Hawking no boundary 
prescription requires us to take the positive sign in the exponent in  Eq.(\ref{HKWKB}),
that is, $\Psi_{+}$. This yields the HH wavefunction. 
Including the $\nu$ term is like solving the wave equation 
with a positive (instead of zero) energy eigenvalue $E=\nu/\lambda^2$
(see Figure (\ref{fig8})), 
\ba
\left( \frac{\hat{\Pi_{a}}^2}{2} + U_{0}(a)\right)\Psi(a) = E \Psi (a)   \nonumber 
\ea
The presence of $E>0$ decreases the tunneling amplitude $\Psi_{+}$.
To see the origin of $E$, consider the presence of a generic field $\chi$, so the WdW equation is
crudely given by
\ba
\label{wd}
\left(-\frac{\hat{\Pi_{a}}^2}{2} - U_{0}(a) + \frac{\Pi_{\chi}^2}{2} + \omega (\chi, a)^{2} \right) \Psi(a, \chi) = 0
\ea
Note that the kinetic terms in Eq.(\ref{wd}) have opposite signs. Let 
\ba
\Psi(a, \chi) = \Sigma_{n} \psi_{n}(a) u_{n}(\chi)
\ea
then the above equation can be separated,
\ba
\label{twoeq}
\left(\frac{\Pi_{\chi}^2}{2} + \omega (\chi, a)^{2} \right) u_{n}(\chi)
= \epsilon_{n}(a)u_{n}(\chi) \\ \nonumber
\left(-\frac{1}{2}\frac{d^2\psi_{n}}{da^2} + U_{0}(a)\psi_{n}\right)
= \epsilon_{n}(a)  \psi_{n} 
\ea
Taking the ground state, we obtain the zero point energy  $\epsilon_{0}$, with $\Psi(a)=\psi_{0}(a)$. Including the zero point energy of all $\chi$ fields then yields $E$, which turns out to be
 independent of $a$. 

If instead, we take $\Psi_{-}$ in Eq.(\ref{HKWKB}), as suggested by 
Linde and Vilenkin \cite{Linde:1984mx} and which is more familiar in quantum mechanics, 
we see that the inclusion of the $\nu$ term enhances tunneling. 
This means interaction with the environment enhances the tunneling amplitude, which is 
counter-intuitive. Furthermore, the highest tunneling probability will go for large $\Lambda$, 
when the semi-classical approximation used here breaks down. 
%To be consistent with ordinary quantum field theory, 
We believe the Hartle-Hawking no boundary prescription is the correct one for the
spontaneous creation of the universe.

As we see clearly now, this $\nu$ term plays the role of radiation.
The deformed $S^4$ has, in a sense, allow the spontaneous creation of a universe
with some radiation in it. In pure gravity, this is simply gravitational radiation.
Due to the presence of such radiation, the cosmic
scale factor $a(\tau)$ does not vanish any more at the poles of $S^4$.
In fact, the decoherence has led to a flattening of the two poles.

We note that,
in theories with dynamical $\lambda$, there will be momentum terms
associated with $\lambda$ in the Wheeler-DeWitt equation. For example,
if $\lambda$ is associated with the potential energy of some scalar
field $\phi$, then there will be momentum term $\hat{\Pi_{\phi}}$ in
the Wheeler-DeWitt equation. However, the inclusion of such a momentum
term for $\lambda$ is not expected to change the boundedness of the
gravitational potential.

For large cosmological constant $\Lambda$, the radiation is suppressed. In this case, one 
may ignore it. However, as $\Lambda$ decreases and the size of the universe grows, radiation 
becomes important. More radiation also means stronger suppression of the tunneling, consistent
with the intuitive picture of tunneling while interacting with the environment. It is this suppression that provides a bound to the Euclidean gravity action.

%To evaluate the above expression further we need to know the number of modes 
%that contribute to the decoherence. There are two natural cut-offs in the 
%problem. The infrared cut-off comes from the Hubble
%length $H^{-1}$. We do not expect larger wavelenths to play any role 
%in the problem. The ultra-violet cut-off comes from the string length 
%$l_{s}$. Since the substcript ``$n$''  really is a shorthand notation
%for ${n,l,m}$, the number of modes between the cutoff scales 
%$H^{-1}$ and $l_{s}$ are :

%\baray
%N = \left ( \frac{H^{-1}}{l_{s}} \right ) ^{3}
%\earay

%As $N$ will be a large number and $\omega_{n}(\alpha)$ will 
%quickly become much larger than $T^{-1}$, therefore, 
%we can approximate Eq.(\ref{dec}) as

%\baray
%\Psi[\alpha] = \int D[\alpha]  
%e^{-I^{E}_{o}[\alpha]} \prod_{n \geq 1} e^{-\omega_{n}(\alpha)T/2}
%\earay
%Doing a steepest descent calculation, this leads to
%\baray
%\Psi[\alpha] = e^{\frac{3\pi}{G\Lambda} - NT <\omega>/2}
%\earay
%where $<\omega>$ is the average frequency given by
%\baray
%<\omega> = \frac{1}{N} \int_{0}^{N} dn \omega(n,\alpha) = \frac{H N}{2}
%\earay

\section{Connection to SOUP}

In \cite{Firouzjahi:2004mx} we lay out the motivation for SOUP, short
for ``Selection of the Original Universe Principle''. This is an alternate
to the Anthropic Principle.  Since observational evidence of an 
inflationary epoch is very strong,
we suggest that the selection of our particular vacuum state 
follows from the evolution of the inflationary epoch.
That is, our particular vacuum site in the cosmic landscape
must be at the end of a road that an 
inflationary universe will naturally follow. Any vacuum state that
cannot be reached by (or connected to) an inflationary stage can be 
ignored in the search of candidate vacua. That is, the issue of the
selection of our vacuum state becomes the question of the selection 
of an inflationary universe, or the selection of an original universe 
that eventually evolves to an inflationary universe, which then 
evolves to our universe today.
The landscape of inflationary states/universes should be much better 
under control, since the inflationary scale is rather close to
the string scale.
In \cite{Firouzjahi:2004mx} we proposed that, by analyzing all 
known string vacua and string inflationary scenarios, one may 
be able to pin down SOUP. Here, the rate 
of tunneling from nothing (i.e., no classical spacetime) to a 
deSitter universe (much like the inflationary universe) is now
given by $\Gamma \simeq \exp(F)= \exp(-S_{E,dC})$, where
\ba
\label{FS1}
- F = S_{E,dC} = S_{E,0}+ D \simeq -\frac{3 \pi}{G\Lambda} + 
\frac{n_{dof}}{4 \pi^2} \frac{V_4}{l_s^4}
\ea
where $D$ is the decoherence term, $l_s$ is the cut-off scale and $n_{dof}$ is the number of light degrees of freedom.  Here $V_4$ is the 4-volume of the 
instanton. For $S^4$, $V_4 = 8 \pi^2 r^4/3= 24 \pi^2/{\Lambda^2}$ is its area.
In string theory, that cut-off is naturally provided by the string scale.
Note that $S_{E,dC}$ is now bounded from below. 

To gain some idea of the magnitude of $F$, we use Eq.(\ref{FS}) to crudely estimate the value of $\Lambda$ with maximum tunneling probability,
\ba
\Lambda_{max} \simeq \frac{4 n_{dof} G}{\pi l_s^4} \nonumber
\ea
so that
\ba
 F_{max} \simeq \frac{3}{2n_{dof}} (2 \pi M_{Pl}  l_s)^4 \nonumber
\ea
For the string scale a few orders of magnitude smaller than $M_{Pl}$, we see that $\Lambda_{max}$ takes a value quite close to that expected in an inflationary universe,
with $F \sim 10^{12}$. On the other hand, the critical value of $\Lambda$ is $\Lambda_c \simeq \Lambda_{max}/2$. For $\Lambda < \Lambda_c$, the barrel-shaped 
instanton is destroyed. At $\Lambda_c$,
$F(\Lambda_c)\simeq 0$, so the tunneling probability at $\Lambda$ close to $\Lambda_c$ is already negligibly small compared to that at $\Lambda_{max}$.
Tunneling to a supersymmetric vacuum is totally suppressed.
For $l_s$ around the $TeV$ scale \cite{Arkani-Hamed:1998rs}, we find that 
$\Lambda_{max}$ is quite close to today's dark energy value. This is similar in spirit to Ref\cite{Hsu:2004jt}. 
However, this scenario will imply that our universe has not really gone through the 
whole hot big bang and certainly not the standard inflationary epoch.

The key tool is the modified bounce. Although we only deal with the 
$S^4$ instanton and its modification 
in this paper, one can easily generalize this analysis to higher dimensions.
We expect the form of the tunneling probability to be 
$P \sim e^F$, with $F = -S_{E,dC}= -S_E -D$. In $10$-D, 
\ba
\label{new10S}
S_{E,dC} \simeq S_{E,10}  + c \left(\frac{V_{10}}{l_s^{10}}\right)
\ea
where $S_{E,10}$ is the 10-D Euclidean action determined in mini-superspace and 
$V_{10}$ is the 10-dimensional volume of the instanton. 
In effective $4$-D theory, $S_{E,10}$ reduces to $-{3 \pi}/{G\Lambda}$,
since 
$$8 \pi G = M_{Pl}^{-2}=\frac{g_s^2 l_s^8}{4 \pi V_6}$$ 
where $V_6$ is the 6-D compactification volume.
(Recall $\alpha^{\prime}=M_s^{-2}=(l_s/2 \pi)^2$.)
The constant $c$ has to be calculated for each vacuum and will
depend on the details of the vacuum.
To get an order of magnitude estimate of $c$ we have, by comparing to the 4-D case,
\ba
c \simeq \frac{n_{dof}}{ \pi} \frac{1}{M_{Pl}^2 l_s^2 g_s^2}
\ea
so we expect $c$ to be small.
In \cite{Firouzjahi:2004mx} we have considered such instantons. We did
not do a careful calculation of $\nu$ then, that has been the main 
content of the present paper, but we phenomenologically guesstimated a 
possible range for its values and calculated the creation probabilities for various
vacua in the cosmic landscape. 
 Among all the known string vacua, we find that the universe 
most likely to be created is a KKLMMT type inflationary vacuum. 
(For details, we refer the reader
to \cite{Firouzjahi:2004mx}.) The probabilities calculated were very
robust and did not depend on the exact value of $c$. 
Changing from $V_9$ to $V_{10}$ does not change the overall qualitative results. 
Therefore, we expect the conclusions to stay the same.

There are some minor differences.
There we find that tunneling to a universe with today's dark energy is very much suppressed.
Here we see that the cosmological constant corresponding to today's dark energy is actually 
below the critical value, so tunneling directly to today's universe is simply zero.
A more careful calculation based on Eq.(\ref{new10S}) is clearly needed.
However, the important point
is that one can calculate it, at least in principle, for any
given vacuum and, therefore, compare their probabilities. This would
be a huge improvement over the Anthropic Principle. \\

 One can calculate the tunnelling probability to $S^{10}$ using 
Eq.(\ref{new10S}). For $S^{10}$, $S_{E,10} = \frac{\Lambda V_{10}}
{32 \pi G_{10}}$, where $V_{10} = \frac{2^{11} 5! \pi^5}{10!}/
{H^{10}}$, is the volume of $S^{10}$. Note that for $S^{10}$,
$H^2 = \Lambda/ 36$. Without branes, the solution we consider is a 10D supersymmetric vacuum.
Tunneling to this vacuum is suppressed. Next we consider tunneling to a deSitter-like vacuum. 
Let there be $N$ $D9$ brane-antibrane pairs
in the $S^{10}$. So 
\ba
\Lambda = 2N \times 8\pi G_{10} \times \frac{M_s^{10}}{(2\pi)^9 g_s}
= \frac{g_s M_s^2 N}{(2\pi)^4}
\ea
Maximizing $F$ in terms of $N$ gives
\ba
N = \frac{5 n_{dof}(2\pi)^6}{g_s}\left( \frac{M_s}{M_{Pl}}\right)^2
\ea
This means $N = 1$ is a reasonable choice for
%$n_{dof} = 2$, 
$g_s \sim 1$, and $M_s/M_{Pl} = 10^{-3}$. One can then calculate $F$ for $N = 1$ for $S^{10}$. 
In this case, $F$ is dominated by $S_{E,0}$. One gets 
\ba
F(S^{10}) \simeq -S_{E,0}(S^{10}) = \frac{2^{11}~ 5!~ (36 \pi)^5}{10!} \simeq 4 \times 10^{9}
\ea

Following Ref\cite{Firouzjahi:2004mx}, we see that other related geometries such as $S^5 \times S^5$, $S^4 \times S^6$, $S^4 \times S^3 \times S^3$ etc. have slightly smaller but similar values of $F$.

For a KKLMMT-like inflationary universe, the choice of fluxes fixes the string scale as well as 
the scale of inflation, where $G$ and the density perturbation measured in the cosmic microwave background radiation are used as input parameters. Using Eq.(\ref{FS1}) and $\Lambda = 8\pi G \rho_{vac}$, where the compactified volume has been absorbed into $G$, 
one can rewrite Eq.(\ref{FS1}) so
\ba
F = \frac{3}{8}\left( \frac{\sqrt{8\pi}M_{Pl}}{\rho_{vac}^{1/4}}\right)^4 - \frac{3}{4 (2\pi)^6} \left( \frac{M_{s}}{\sqrt{8\pi}M_{Pl}}\right)^4 \left( \frac{\sqrt{8\pi}M_{Pl}}{\rho_{vac}^{1/4}}\right)^8
\ea
Plugging in the above mentioned values for the mass scales, we get $F$. 
For a typical choice of fluxes that would give $M_s \sim 10^{15}$GeV and an inflationary scale $\rho_{vac}^{1/4} \sim 10^{14}$GeV, with the Planck mass $M_{Pl} \sim 10^{18}$GeV, one gets
 $$F \sim 10^{16}$$
 for a realistic brane inflationary scenario.
This big increase in $F$ from the 10D scenario is a result of the decrease in the effective cosmological constant due to the warping of the geometry. By varying the RR and the NS-NS fluxes, one can maximize $F$ in a way which is not possible in 10D.  Of course, this value is sensitive to the details of the model. A more careful calculation will be important.

\section{Discussion}

Let us make some comments here.

\itemize

\item The loop correction to the Euclidean action has been calculated before
\cite{Hawking:1976ja,Gibbons:1978ji}. It has the form
$\log(\Lambda/\mu)$, so $S_E$ becomes 
$$S_E= - \frac{3 \pi}{G\Lambda}  \to  - \frac{3 \pi}{G\Lambda}\left(1 + \beta G\Lambda\log(\Lambda/\mu)\right)$$
where $\beta$ measures the number of fields involved.
Ref\cite{Barvinsky:1992dz} considers the metric perturbation also; however, 
they then relate it to the one-loop contribution, yielding the above result.
Note that this loop correction does not solve the boundedness problem in the
HH wavefunction. Its implications on probability of inflation has also been
discussed (see \cite{Barvinsky:1996ce} and the references therein). 
 
\item Decoherence effects are to be distinguished from the 
particle creation effects that have been discussed in the literature
\cite{Rubakov:1984bh}.
Particle creation effects arise at the one-loop level, while the
decoherence effect discussed here is a back-reaction improved quantum correction. 

\item Back reaction is crucial in obtaining the decoherence term. This reminds one of the 
situation in quantum field theory, where
a simple 1-loop correction to the coupling takes on new significance when it is 
renormalization group improved. 

\item Integrating out the environment provides a new interaction term for
the cosmic scale factor $a$ in the Einstein equation. This decoherence term 
behaves just like ordinary radiation and generates an effective term in the Euclidean action for the instanton. 
The presence of this term provides the lower bound to the resulting Euclidean gravity.

%\item It is interesting to note that in terms of the string coupling $g_s$, an
%O$({\nu}/{\lambda^2})$ correction to the Euclidean
%action corresponds to an order O$({g_s^{2}})$ correction.
%This is a novel feature. ?

\item The correction to the Hartle-Hawking wavefunction 
results in a decrease in the Euclidean action. If we are
to interpret the Euclidean action as the entropy, then
there is a decrease in the entropy due to decoherence.
This is in accordance with the Bekenstein bound on entropy.
A pure deSitter space provides the upper bound on entropy.
Decoherence, which is due to the inclusion of extra degrees
of freedom, then leads to a lower entropy respecting the entropy bound.

\item Once the inflationary universe is created, inflation simply red-shifts 
the radiation so the radiation term quickly becomes negligible.  
It seems that the initial radiation is present just to suppress the tunneling.

\item It is interesting to speculate what happens after inflation.
Towards the end of inflation, the inflaton field (or the associated
tachyon mode) rolls down to the bottom of the inflaton potential. The
universe is expected to be heated up to start the hot big bang epoch.
It is not unreasonable to expect the wavefunction of the universe to be
a linear superposition of many vacua (say, a billion of them) at the
foothill of the inflaton potential. This is like a Bloch wavefunction
as proposed in Ref\cite{Kane:2004ct}. Since tunneling happens only
between vacua with (semi-)positive cosmological constants, this
wavefunction should be a superposition of vacua with only positive
cosmological constants. The ground state energy of this wavefunction is
expected to be much smaller than the average vacuum energy, a property
of Bloch wavefunctions. As the universe evolves, tunneling among the
vacua will become suppressed, due to cooling as well as decoherence.
Eventually, the environment collapses the wavefunction to a single
vacuum. Since the Bloch wavefunction is able to sample many vacua, it
is natural to expect that it will collapse to the vacuum state with the
smallest positive vacuum energy within teh sample. This may partly
explain why the dark energy is so small.

\item In terms of recent work on the statistics of the landscape 
\cite{Kachru:2003aw,Douglas:2003um},
one may view our work as providing a measure to the counting
of vacua. Each vacuum is weighted with its probability of tunneling 
from nothing. Since the exponent of the tunneling probability can 
differ by many orders of magnitude, this is likely to be the dominant 
contribution to the measure. We argue that a 4-dimensional inflationary 
universe very much like ours has large probability; this implies that vacua
that cannot be reached after inflation should have vanishing measure.

\item The effect of decoherence on other tunneling problems in quantum gravity
should be re-examined. These include tunneling in eternal inflation, the 
Coleman-deLuccia tunneling etc.

\section{Summary and Remarks}
     
We address a number of questions in this paper.

First we ask what happens to the well known
$S^4$ bounce solution to the Euclidean Einstein equations 
when the perturbations to the metric are taken into account. 
How does the bounce change as a result?
In particular, is there a parameter
characterizing the perturbations that describes the modification
of the bounce? Is there a range of this parameter that destroys 
the bounce altogether? 

To answer the first question, we apply the path integral
techniques to trace out the perturbations. The result is summarized in the introduction.
Taking back-reaction into account, the Euclidean action now includes an additional term that encapsulates
the effect of the perturbations. Depending on the value of the 
parameter $c$ or equivalently $\nu$, the bounce gets deformed. For a fairly
large range of this parameter, the effect is to just deform the bounce
solution. Since this contribution is always positive, the tunneling
is always suppressed. At a critical value, the bounce is destroyed. One 
may interpret that tunneling is forbidden in this limit. 

Next, we ask what these perturbations do to the
boundedness of the Euclidean gravitational action. As is well known,
the Euclidean gravitational action for a closed spacetime is not
bounded from below for theories with a dynamical cosmological constant. 
This manifests, for example, as the infinite peaking of
the wavefunction of the universe at the vanishing value of $\Lambda$.  
Does the inclusion of the perturbations change anything here? 

To answer this question, we see that the effect of the perturbations is
to change the wavefunction of the universe from $\exp (3\pi/G\Lambda)$
to $\exp (3\pi/G\Lambda - C/\Lambda^2)$ where $c$ depends on the 
particular perturbations that we are looking at. So, at least for
the problem at hand, the inclusion of the perturbations and their back reaction makes the
Euclidean action bounded from below. This renders the wavefunction 
normalizable. In the study of tunneling, it makes no sense
to talk about gravity without taking into account the perturbations 
to the metric. In this sense, quantum gravity is consistent as long as 
we are careful to include the effects of the perturbative modes which are 
always present.

Once we have a sensible wavefunction, we can now go ahead and 
apply it to the cosmic landscape in string theory. Since Euclidean action and 
tunneling probability are both dimensionless, we can compare the tunneling 
from nothing to any point in the landscape and find the sites that have the largest 
probability. This program hopefully will allow us to understand why we are 
where we are, without resorting to the anthropic principle.

\vspace{0.5cm}

{\large{\bf{Acknowledgments}}}\\

We thank Faisal Ahmad, Andrei Barvinski, Spencer Chang, Jacques Distler, 
Hassan Firouzjahi, Lerrain Friedel, Jim Hartle, Gordy Kane, Louis Leblond, Andrei Linde, 
Juan Maldacena, Gautam Mandal, Liam McAllister, Lubos Motl, 
Hirosi Ooguri, Koenraad Schalm,  Jan Pieter van der Schaar, Jim Sethna,  Sarah Shandera, Gary Shiu, Ben Shlaer
and  Cumrun Vafa for useful discussions. This work is supported by the National Science Foundation under Grant No. PHY-009831.

\vspace{0.5cm}

\appendix

\section{A Quantum Mechanical Example}

To see the basic idea, recall the quantum tunneling of the system
\ba
L_0= \frac{M}{2} {\dot q}^2 -V(q)
\ea
with a quartic $V(q)$ as shown in Fig. 2. The tunneling rate 
$\Gamma$ from the local minimum at $q=0$ to the exit point 
$q=q_0$ is well-known,
\ba
\Gamma &=& A \exp (-S_0) \nonumber \\
S_0 &=& \int^{q_0} \sqrt{2MV(q)} dq =\int  \left 
(\frac{M}{2} {\dot q}^2 +V(q) \right) d\tau
\ea
where $\tau$ is the Euclidean time and $S_0$ is the bounce, 
i.e., the instanton solution \cite{Coleman:1977py}.
This WKB approximation is good provided that the height of 
the barrier is larger than $\om_0$, where 
$$M \om_0^2 =  \frac{\partial^2 V}{\partial q^2}|_{q=0}$$
Note that, for $V(q)$ bounded from below, $S_0$ is bounded 
from below, as required by consistency.
In a more realistic situation, the particle interacts with 
the environment. Typically, this introduces a frictional force, 
so the corresponding classical equation is given by
\ba
M{\ddot q} + \eta {\dot q} + \frac{\partial V}{\partial q} = 0
\ea
The impact of such a frictional term on the particle is to 
suppress the tunneling rate. Consider the following system
\ba
L= \frac{M}{2} {\dot q}^2 -V(q) + \frac{1}{2} \sum_{\al} m_{\al} 
\left({\dot x_{\al}}^2 - \om_{\al}^2 x_{\al}^2\right)
-q \sum_{\al} C_{\al}x_{\al} -\frac{1}{2}M (\delta \om)^2 q^2
\ea
where one may consider $q$ to be the system and the $x_{\al}$ to 
be the environment. The last term is a counter term introduced to 
correct the shift in frequency,
\ba 
M (\delta \om)^2 &=&  \sum_{\al} \frac{C^2_{\al}}{m_{\al} \om^2_{\al}} 
\ea
The interactions of $q$ with the $x_{\al}$ introduces the friction term
\ba
\eta = \frac{\pi}{2} \sum_{\al} \frac{C_{\al}^2}{m_{\al} \om_{\al}^2} 
\delta(\om_{\al }-\om) 
\ea
for $\om$ smaller than some critical $\om_c$.

The tunneling rate of this sytem can be easily found \cite{Sethna:1981dr,Caldeira:1982uj} 
that the bounce $S_0$ increases to
\ba
S \simeq S_0 \left( 1 + \frac{\eta}{2M \om_0}\right)
\ea
That is, the interaction with the environment, or the friction, 
suppresses the tunneling rate. This qualitative feature remains 
true when the environment and its interaction with the
system is more complicated.

\section{Calculations on the Squashed $S^4$}

This appendix gives details for Section $3$. We show how
we perform the path integral over the $t_n$ mode.
The solution to Eq.(\ref{t_n}) satisfying the initial condition 
$t_n (x = 1 - \delta) = t_n^i$, and the final condition 
$t_n (x = -1 + \delta) = t_n^f$,  is given by
\ba
\label{tnx}
t_n(x) = \frac{F_n(x)}{\sqrt{1 - x^2}} 
= \frac{A P^{-n}_{1}(x) + B Q^{n}_{1}(x)}{\sqrt{1 - x^2}}
\ea 
where $A$ and $B$ are given by
\ba
\label{AB}
A = \sqrt{2\delta - \delta^2}\frac{t_n^i Q^n_{1}(-1 + \delta) - t_n^f 
Q^n_{1}(1 - \delta)}{ P^{-n}_1(1 - \delta) Q^n_{1}(-1 + \delta) - 
 P^{-n}_1(-1 + \delta) Q^n_{1}(1 - \delta)} \\ \nonumber
B = - \sqrt{2\delta - \delta^2}\frac{t_n^i P^{-n}_{1}(-1 + \delta) - t_n^f 
P^{-n}_{1}(1 - \delta)}{ P^{-n}_1(1 - \delta) Q^n_{1}(-1 + \delta) - 
 P^{-n}_1(-1 + \delta) Q^n_{1}(1 - \delta)}
\ea

The trace operation is defined as doing the following
\ba
\int dt_n^i \int dt_n^f \delta (t_n^i - t_n^f)\int^{(t_n^{initial} 
= t_n^i)}_ {(t_n^{final} = t_n^f)}
D[t_n] \exp \left( S^n_E [t_n(x)]\right)
\ea

Since our purpose finally is to take to trace over the $t_n$ mode, we can
set $t_n^i = t_n^f$ at this stage. The Euclidean action due to the $n$th 
mode is then
\ba
S_{E}^{n} & = & \frac{1}{2} \left[a(\tau)^3 t_n \frac{dt_n}{d\tau}  
\right]_{x = (1 - \delta)}^{x = (-1 + \delta)} \\ \nonumber
& = & \frac{1}{2} \left[xF_n^2 + (1 -x^2)F_n\frac{dF_n}{dx}  
\right]_{x = (1 - \delta)}^{x = (-1 + \delta)}
\ea

Using Eq.(\ref{tnx}, \ref{AB}) and the symmetry properties of
the associate Legendre functions
\ba
P^{-n}_1(-x) = - P^{-n}_1(x), &  \quad & Q^{n}_1(-x) = Q^{n}_1(x) \\ \nonumber
\partial_{x}P^{-n}_1(-x) = \partial_{x}P^{-n}_1(-x), & \quad &
\partial_{x}Q^{n}_1(-x) = - \partial_{x}Q^{n}_1(x)
\ea
one gets the following action
\ba
S_{E}^{n} = \delta(2 -\delta) \left[ (1 - \delta) + \delta(2 -\delta)
\frac{\partial_{x}Q^{n}_1(1 -\delta)}{Q^{n}_1(1 -\delta)}\right] (t_n^i)^2
\ea
Next we must find the prefactor to the path integral for $t_n$.
This can be found as explained in the Appendix C. The prefactor 
is given by $\frac{1}{\sqrt{2\pi g_n(-1 + \delta)}}$, where $g_n(x)$
is a solution to Eq.(\ref{t_n}), i.e.
$ g_n(x)= \frac{C P^{-n}_{1}(x) + D Q^{n}_{1}(x)}{\sqrt{1 - x^2}}$, 
and it satisfies the following conditions
\ba
g_n(1 -\delta) = 0, \quad \partial_x g_n(1 -\delta) = 1
\ea
It is easy to check that such a solution to Eq.(\ref{t_n}) 
has the following values
of $C$ and $D$
\ba
C = \sqrt{2\delta - \delta^2} \frac{Q^n_1(1-\delta)}{ Q^n_{1}(1 - \delta)
\partial_x P^{-n}_1(1 - \delta) 
 - P^{-n}_1(1 - \delta) \partial_x Q^n_{1}(1 - \delta)} \\ \nonumber
D = - \sqrt{2\delta - \delta^2} \frac{P^{-n}_1(1-\delta)}{ Q^n_{1}(1 - \delta)
\partial_x P^{-n}_1(1 - \delta) 
 - P^{-n}_1(1 - \delta) \partial_x Q^n_{1}(1 - \delta)}
\ea
The prefactor is then given by
\ba
\frac{1}{\sqrt{2\pi g_n(-1 + \delta)}} = \frac{1}{\sqrt{4\pi}}
\sqrt{\left( \frac{\partial_xQ^n_1(1-\delta)}{Q^n_1(1-\delta)} 
- \frac{\partial_xP^{-n}_1(1-\delta)}{P^{-n}_1(1-\delta)} \right)}
\\ \nonumber
= \frac{1}{\sqrt{4\pi}} \frac{\sqrt{Q^n_1(1 -\delta) P^n_1(1 - \delta) (2\delta- \delta^2)}}{2^n} \sqrt{\frac{\Gamma\left( \frac{2-n}{2}\right) \Gamma\left( \frac{3-n}{2}\right)}{\Gamma\left( \frac{2+ n}{2}\right) \Gamma\left( \frac{3+n}{2}\right)} }
\ea

\section{Path Integral of an Oscillator with variable 
mass and variable frequency}

Here we review some basic properties of path integral and apply them to the evaluation of the 
prefactor in Eq.(\ref{PII}). To be concrete, we shall follow Ref\cite{Felsager:1981iy}.
Consider an oscillator with mass $m(t)$ and frequency $\omega(t)$. What
we have in mind, in particular, is a case like Eq.(\ref{PI}) which
describes the path integral for a tensor mode (in Euclidean spacetime).
Comparing Eq.(\ref{PI}) with this appendix, the time dependent mass
would be $a(t)^3$ and the time dependent frequency would be
${(n^2 -1)}/{a^2}$ .  
We would like to evaluate the propagator $K(x_f,T | x_i, 0)$. The action
is given by
\baray
& S = \int_{o}^{T} dt \left( \frac{1}{2} m \left( \frac{dX}{dt}\right)^2 
- \frac{1}{2} m \omega(t)^2 X^2 \right) 
\earay
The problem can be simplified by introducing a new "time" variable $u$ so as to map the 
present problem to that of an oscillator with unit mass and variable frequency: 
\baray
\label{defnu}
du = \frac{dt}{m(t)}
\earay
In terms of $u$ the action becomes
\baray
& S = \frac{1}{2} \int_{u_i}^{u_f} du \left( \left( \frac{dX}{du}\right)^2 
- \Omega(u)^2 X^2 \right)
\earay
that is, the action for an oscillator with unit mass and variable 
frequency $\Omega(u)$. As this is quadratic in $X(u)$, we can 
expand around a classical solution, $X = x_{cl} + x$, where the classical 
solution $x_{cl}(u)$ satisfies the equation of motion
\ba
\label{class_eqn}
\frac{d^2 x_{cl}(u)}{du^2} + \Omega(u)^2 x_{cl}(u) = 0
\ea
where $\Omega(u)$ depends on $n$. For example,
for the $n$th tensor field perturbation, $\Omega(u) = (n^2 -1)a(u)^4$. 
One can use the above solution for $x_{cl}$, where $x_{cl}(u_i)=X(u_i)$ and 
$x_{cl}(u_f)=X(u_f)$, to calculate the classical action $S[x_{cl}]$. This gives the
saddle point value of the path integral. Following the discussion in Sec. 5, 
the prefactor of the propagator is given by:
\baray
\label{prefactor}
& \int_{x(u_i)=0}^{x(u_f)=0}D[x(t)] \exp \left[ \frac{i}{2} \int_{u_i}^{u_f}du
\left(\left( \frac{dx}{du}\right)^2 - \Omega(u)^2 x^2  \right) \right] 
\earay

By a further change of variables, we can change the action to a free
particle action. To do so, let $g(u)$ be a solution of the 
equation of motion (\ref{class_eqn})
%\baray
%\label{defng}
%\frac{d^2 g(u)}{dt^2} + \Omega(u)^2 g(u) = 0
%\earay
such that $g(u)$ does not vanish at the initial end point
\baray
g(u_i) \neq 0
\earay
implying that $g(u)$ is not a path in Eq.(\ref{prefactor}).
Define the following transformation of variables
\baray
\label{defny}
 x(u) = g(u) \int_{u_i}^{u}\frac{y'(s)}{g(s)}ds=g(u)F(u)
\earay
where the function $y(u)$ obeys $y(u_i) = 0$. 
Differentiating Eq.(\ref{defny}), one finds the inverse transformation
\baray
& y(u) = x(u) - \int_{u_i}^{u}\frac{g'(s)}{g(s)}x(s)ds
\earay 
In terms of the $y$ variable, we have:
\baray
& \left( \frac{d^2}{du^2} + \Omega(u)^2 \right)x(u) = \left(g''(u) 
+ \Omega(u)^2 g(u) \right) \int_{u_i}^{u}\left[ \frac{y'(s)}
{g(s)}\right]ds \\ \nonumber
& + \frac{g'(u)y'(u)}{g(u)} + y''(u) = \frac{g'(u)y'(u)}{g(u)} +   y''(u)
\earay
%But here the first term on the right vanishes due to Eq.(\ref{defng}). So,
so the action becomes
\baray
& S[x(u)] = -\frac{1}{2} \int_{u_i}^{u_f}du \left[ F(u)g'(u)y'(u)+ 
F(u)g(u)y''(u)\right]
\earay
%with $F(u) = \int_{u_i}^{u}ds\frac{y'(s)}{g(s)}$. 
Keeping in mind that $x(u)$ vanishes at the boundaries (see Eq.(\ref{prefactor})), 
a further partial integration then leads to
\baray
& S[x(t)] = \frac{1}{2}\int_{u_i}^{u_f}du \left[ \frac{dy}{du} \right]^2
\earay
This is just a free particle action, with boundary conditions
%However the boundary condition for this free particle at $u = u_f$ is nonlocal
\baray
& y(u_i) = 0 ~~ ; \quad & y(u_f) = \int_{u_i}^{u_f}ds \frac{y'(s)}{g(s)} = 0
\earay
To impose the (non-local) boundary condition $x(u_f) = y(u_f)=0$, we introduce
\baray
& \delta(x(u_f)) = \frac{1}{2\pi} \int d\alpha ~ e^{-i \alpha x(u_f)}
\earay
The path integral can now be written as
\baray
&\int_{x(u_i)=0}^{x(u_f)=0} D[x(u)]\exp \left(iS[x(u)] \right)   
= \frac{1}{2\pi} \int_{x(u_i)=0}^{x(u_f)~arbitrary} \int_{-\infty}^{\infty}
d\alpha  D[y(u)] \nonumber \\  & \det \left[ \frac{\delta x}{\delta y}\right].
\exp \left[-i \alpha g(u_f)\int_{u_i}^{u_f}ds 
\frac{y'(s)}{g(s)} \right]. \exp \left[  \frac{i}{2}\int_{u_i}^{u_f}du
\left( \frac{dy}{du}\right)^2\right]  
\earay

As the transformation between $x(u)$ and $y(u)$ is linear, the Jacobian
$\det \left[ \frac{\delta x}{\delta y}\right]$ is independent of $y(u)$.
To carry out the above path integral, we just have to complete the 
square by the use of the new variable
\baray
& \gamma(u) = y(u) - \alpha ug(u_f) \int_{u_i}^{u}\frac{ds}{g(s)}
\earay
The above path integral thus becomes
\baray
 ... &= \frac{1}{2\pi} \det \left[ \frac{\delta x}{\delta y}\right] 
\int_{-\infty}^{\infty} d\alpha \exp \left[ -\frac{i}{2}\alpha^2 
g(u_f)^2 \int_{u_i}^{u_f}\frac{du}{g(u)^2} \right] \\ \nonumber 
& \int_{\gamma(u_i)=0}^
{\gamma(u_f)arbitrary} D[\gamma(u)] \exp \left[ \frac{i}{2}
\int_{u_i}^{u_f} \left( \frac{d\gamma}{du}\right)^2 \right]
\earay

It is easy to carry out the $\alpha$ integral. To do the remaining 
path integral, one just has to notice that it represents the 
probability amplitude for finding the particle
{\it anywhere} at the  time $u_f$. This probability is obviously $1$.
\baray
& \int_{\gamma(u_i)=0}^
{\gamma(u_f)} D[\gamma(u)] \exp \left[ \frac{i}{2}
\int_{u_i}^{u_f} \left( \frac{d\gamma}{du}\right)^2 \right]
\\ \nonumber & = \int_{-\infty}^{\infty} dx K(x, u_f | 0, u_i) = 1 
\earay
where $\gamma(u_f)$ is arbitrary. Therefore, we have 
\baray
& \int_{x(u_i)=0}^{x(u_f)=0}D[x(t)] \exp \left[ \frac{i}{2} \int_{u_i}^{u_f}du
\left(\left( \frac{dx}{du}\right)^2 - \Omega(u)^2 x^2  \right) \right] \\ \nonumber
& = \det \left[ \frac{\delta x}{\delta y}\right] \sqrt{\frac{1}{2\pi 
i g(u_f)^2 \int_{u_i}^{u_f}\frac{ds}{g(s)^2}}}
\earay

It remains to find the value of the Jacobian. It can be found by a discretization
process. The paths $x(u)$ and $g(u)$ can be replaced by the multidimensional points
$(x_0, x_1, ..., x_N)$ and $(y_0, y_1,..., y_N)$ with $x_k = x(u_k)$ and
$y_k = y(u_k)$. The linear transformation can then be approximated as
\baray
y_n & = & x_n - \frac{T}{N}\sum_{k=1}^{n}\frac{g'(u_k)}{g(u_k)} 
\frac{x_k + x_{k-1}}{2} 
\earay
Then,
\baray
 J_N = \det \left[ \frac{\partial y_i}{\partial x_j} \right] 
 = \prod_{k=1}^{N}\left( 1 - \frac{1}{2} \frac{g'(u_k)}{g(u_k) \frac{T}{N}} \right)
\earay
Taking the $N \to \infty$ limit, one gets
\baray
 \det \left[\frac{\delta y}{\delta x}\right] =  \exp \left[ -\frac{1}{2}
\int_{u_i}^{u_f}du \frac{g'(u)}{g(u)} \right]  = 
\sqrt{\frac{g(u_i)}{g(u_f)}} 
\earay
So the path integral prefactor becomes
\baray
\label{pref1}
& \int_{x(u_i)=0}^{x(u_f)=0}D[x(t)] \exp \left[ \frac{i}{2} \int_{u_i}^{u_f}du
\left(\left( \frac{dx}{du}\right)^2 - \Omega(u)^2 x^2  \right) \right] \\ \nonumber
&=\Delta \left[\det \left( - \frac{d^2}{du^2} - \Omega(u)^2  \right) 
\right]^{-1/2}  = \left({2\pi i g(u_i) g(u_f) \int_{u_i}^{u_f}\frac{ds}{g(s)^2}} \right)^{-1/2} 
\earay
where $\Delta$ is a normalization factor.
One can check the validity of this result for simple cases like free particle
and simple harmonic oscillator. 
%We can simplify this result further. We can rewrite the path integral as:
%\baray
%& \int_{x(u_i)=0}^{x(u_f)=0}D[x(t)] \exp \left[ \frac{i}{2} \int_{u_i}^{u_f}du
%x(u)\left(- \frac{d^2}{du^2} - \Omega(u)^2 \right)x(u) \right] \\ \nonumber
%& = \Delta \left[\det \left( - \frac{d^2}{du^2} - \Omega(u)^2  \right) 
%\right]^{-1/2} 
%\earay
Using the result from Eq.(\ref{pref1}), we can write
\baray
\label{ratio}
\frac{\left[\det \left( - \frac{d^2}{du^2} - \Omega(u)^2  \right) 
\right]}{\left[\det \left( - \frac{d^2}{du^2} - U(u)  \right) 
\right]} = \frac{g(u_i) g(u_f) \int_{u_i}^{u_f}\frac{ds}
{g(s)^2}}{G(u_i) G(u_f) \int_{u_i}^{u_f}\frac{ds}
{G(s)^2}}
\earay
where $G(u)$, analogous to $g(u)$, is a solution of the equation
of motion with an arbitrary $U(u)$. And, like $g(u)$, 
$G(u)$ also does not vanish at $u_i$. Now, let $g^0$ denote the
unique solution to $ { \partial_{u}^2 + \Omega(u)^2}g(u) = 0 $
which satisfies the boundary conditions:
\baray
\label{go}
& g^0(u_i) = 0; \quad  & \frac{d}{du}g^0(u_i) = 1
\earay
Let $g^1$ represent the solution that satisfies
\baray
& g^1(u_i)= 1 ;\quad  & \frac{d}{du}g^1(u_i) = 0
\earay
In Eq.(\ref{ratio}), let
\baray
& g(u) = g^0(u) + \epsilon g^1(u) ; \quad  &   G(u) = G^0(u) + \epsilon G^1(u)
\earay
The integral $\int_{u_i}^{u_f}\frac{du}{[g(t)]^2}$ diverges due to
the vanishing of $g^0$ at $u_i$. However, since almost all the contribution
comes from an infinitely small neighborhood of $u_i$ (in the limit
when $\epsilon \to 0$), it follows that it diverges like
\baray
\int_{u_i}^{u_f}\frac{du}{[u - u_i]^2}
\earay
Consequently,
\baray
& {\stackrel{\lim}{\epsilon \to 0}} \quad
{\int_{u_i}^{u_f}\frac{du}{g(u)^2}}=
{ \int_{u_i}^{u_f}\frac{du}{G(u)^2}} 
\earay
So Eq.(\ref{ratio}) becomes
\baray
\frac{\left[\det \left( - \frac{d^2}{du^2} - \Omega(u)^2  \right) 
\right]}{\left[\det \left( - \frac{d^2}{du^2} - U(u)  \right) 
\right]} = \frac{g^0(u_f)}{G^0(u_f)}
\earay
Taking $U(u) = 0$ (for a free particle), we get
\baray
\frac{\left[\det \left( - \frac{d^2}{du^2} - \Omega(u)^2  \right) 
\right]}{\left[\det \left( - \frac{d^2}{du^2}\right) 
\right]} = \frac{g^0(u_f)}{u_f - u_i}
\earay

Now we can calculate the propagator,
\baray
\label{final}
& K(x_f, T| x_i, 0)  \equiv K(x_f, u_f | x_i, u_i)  = K(0, u_f| 0, u_i)
\exp \left( i S[x_{cl}] \right)  \nonumber \\
& = \left(\frac{\det \left( - \frac{d^2}{du^2} - \Omega(u)^2  \right)} 
{\det \left( - \frac{d^2}{du^2}\right)}\right)^{-1/2} 
K_{0}(0, u_f| 0, u_i) \exp \left( i S[x_{cl}] \right)
 \nonumber \\
& = \sqrt{\frac{1}{2 \pi i g^0(u_f)}} \exp \left( i S[x_{cl}] \right) 
\earay

One can switch back to the $t$ variable now using Eq.(\ref{defnu}). 
$S[x_{cl}]$ is the action for the solution $x_{cl}$ of the classical
equation of motion satisfying the boundary conditions $x(u_i) = x_i$
and $x(u_f) = x_f$.

Now we specialize this calculation of the path integral for an oscillator
with varying mass and frequency to our case. Eq.(\ref{class_eqn}) generally
does not lend itself to simple analysis. However, since we expect the 
higher modes (with large values of $n$) to be the major contributors
to the suppression of the tunnelling, one can use the WKB analysis
to find approximate solutions. This is because, firstly, Eq.(\ref{class_eqn}) 
has the form of a Schrodinger equation for an
arbitrary potential $\Omega(u)^2 $, and, secondly, the variation
in $\Omega(u)^2 x_{cl}(u)$ for higher modes is slow enough. Typically, for
the $n$th mode, $\Omega^2 \sim n^2$ and $\dot{\Omega} \sim n$. 
So $\dot{\Omega} << \Omega^2$.
Using the WKB approximation, we get:
\baray
x_{cl}(u) = \frac{1}{\sqrt{\Omega}} \exp \left(\pm i
\int_{u_i}^{u_f}du \Omega(u)  \right)
\earay

Note that in the Euclidean case, $u \to -i u$ (or, equivalently,
$t \to -i \tau$). This is what we have done in Eq.(\ref{class_soln}, 
\ref{gensol}) to get the solution for the tensor mode $t_{cl}(\tau)$. Using these
classical solutions we have calculated the classical action in
Eq.(\ref{S_E}). 
%In Eq.(\ref{pre_fac}) we have calculated the prefactor. 
As explained in Eq.(\ref{final}), the prefactor is
proportional to $({2 \pi i g^0(u_f)})^{-1/2}$  where
$g^0(u)$ satisfies the boundary conditions given in Eq.(\ref{go}).
Following Eq.(\ref{gensol}) it is clear that for tensor modes,
the two linearly independent solutions to the classical equation of motion 
are $\frac{1}{\sqrt{\omega_n}} \exp \left(\pm \int^u du' \omega_n(u') \right)$
. And the following linear combination satisfies the required boundary 
conditions
\ba
g^0(u_f) = \frac{1}{\sqrt{2\omega_n(u_i)\omega_n(u_f)}}\left( \exp(D_n) 
- \exp(-D_n) \right).
\ea
as this choice of $g^0(u)$ vanishes at $u = u_i$ where $D_n = 0$.
$\frac{dg^0(u)}{du} = 1$ at $u = u_i$.
This yields the result in Eq.(\ref{pre_fac}).

\end{document}